\documentstyle[12pt,aaspp4,epsf]{article}
\input epsf

\def\kms{{\rm km/s}}
\def\cc{{\rm cm ^{-3}}}
\def\mum{\mu {\rm m}}

\def\deg{^\circ}

\def\lsun{{\,L_\odot}}
\def\g0{{\,{\rm G}_0}}

\def\llcii{{\rm L_{\rm [CII]}}}
\def\ffir{{\rm F_{\rm FIR}}}
\def\lfir{{\rm L_{\rm FIR}}}
\def\rat{{\rm L_{\rm [CII]}}/{\rm L_{\rm FIR}}}
\def\2rat{{\rm L_{\rm [CII]}}/{\rm L_{\rm TIR}}}
\def\orat{{\rm L_{\rm [OI]}}/{\rm L_{\rm FIR}}}

\def\lb{{\rm L_{\rm B}}}
\def\6f{{\rm F_\nu(60 \ \mum)}}
\def\f100{{\rm F_\nu(100 \ \mum)}}
\def\frat{\6f/\f100}
\def\i6{{\rm L(60 \ \mum)}}
\def\l1{{\rm L(100 \ \mum)}}
\def\r61{{\6f/\f100}}
\def\lo{{\rm L_{\rm [OI]}}}
\def\rcof{(\llcii+\lo)/\lfir}
\def\ln{\rm L_{\rm [NII]}}
\def\nf{\ln/\lfir}
\def\cn{\llcii/\ln}
\def\brat{\lfir/\lb}
\def\ocrat{\lo/\llcii}
\def\o3f{\rm L_{\rm [OIII]}/\lfir}
\def\tnm{\tablenotemark}

\begin{document}

\title{Far Infrared Spectroscopy of Normal Galaxies:
Physical Conditions in the Interstellar Medium}
\author{S. Malhotra \altaffilmark{1,2},
M. J. Kaufman\altaffilmark{3,4}, 
D. Hollenbach\altaffilmark{4}, 
G. Helou\altaffilmark{5}, 
R. H. Rubin\altaffilmark{4},
J. Brauher\altaffilmark{5},  
D. Dale\altaffilmark{5},  
N. Y. Lu\altaffilmark{5}, 
S. Lord\altaffilmark{5}, 
G. Stacey\altaffilmark{6}, 
A. Contursi\altaffilmark{5}, 
D. A. Hunter\altaffilmark{7}, 
H. Dinerstein\altaffilmark{8} } 

\begin{abstract}

The most important cooling lines of the neutral interstellar medium
(ISM) lie in the far-infrared (FIR).  We present measurements by the
Infrared Space Observatory Long Wavelength Spectrometer of seven lines
from neutral and ionized ISM of 60 normal, star-forming
galaxies. The galaxy sample spans a range in properties such as
morphology, FIR colors (indicating dust temperature), and FIR/Blue
ratios (indicating star-formation activity and optical depth). 

In two-thirds of the galaxies in this sample, the [\ion{C}{2}] line flux is
proportional to FIR dust continuum. The other one-third show a smooth
decline in $\rat$ with increasing $\r61$ and $\lfir/\lb$, spanning a
range of a factor of more than 50. Two galaxies, at the warm and active
extreme of the range have $\rat < 2 \times 10^{-4} (3\sigma$ upper
limit).  This is due to increased positive grain charge in the warmer
and more active galaxies, which leads to less efficient heating by
photoelectrons from dust grains.

The ratio of the two principal photodissociation region (PDR) cooling
lines $\ocrat$ shows a tight correlation with $\r61$, indicating that
both gas and dust temperatures increase together.  We derive a
theoretical scaling between [\ion{N}{2}](122 $\mum$) and [\ion{C}{2}] from ionized
gas and use it to separate [\ion{C}{2}] emission from neutral PDRs and
ionized gas.  Comparison of PDR models of Kaufman et al. (1999) with
observed ratios of (a) $\lo/\llcii$ and $(\llcii+\lo)/\lfir$ and (b)
$\orat$ and $\r61$ yields far-UV flux $G_0$ and gas density $n$. The
$G_0$ and $n$ values estimated from the two methods agree to better
than a factor of 2 and 1.5 respectively in more than half the sources.

The derived $G_0$ and $n$ correlate with each other, and $G_0$
increases with $n$ as $G_0 \propto n^{\alpha}$, where $\alpha\approx
1.4$ . We interpret this correlation as arising from Str\"{o}mgren
sphere scalings if much of the line and continuum luminosity arises
near star-forming regions.  The high values of PDR surface temperature
($270-900\,\rm K$) and pressure ($6\times 10^4-1.5\times 10^7\,\rm
K\,cm^{-3}$) derived also support the view that a significant part of
grain and gas heating in the galaxies occurs very close to
star-forming regions. The differences in $G_0$ and $n$ from galaxy to
galaxy may be due to differences in the physical properties of the
star-forming clouds. Galaxies with higher $G_0$ and $n$ have larger
and/or denser star-forming clouds.

\end{abstract}

\altaffiltext{1}{Johns Hopkins University, Charles and 34th Street, Bloomberg Center, Baltimore, MD 21210}
\altaffiltext{2}{Hubble Fellow}
\altaffiltext{3}{Dept. of Physics, San Jos\'e State University, One Washington Square, San Jos\'e, CA 95192-0106} 
\altaffiltext{4}{NASA/Ames Research Center, MS 245-3, Moffett Field, CA 94035}
\altaffiltext{5}{IPAC, 100-22, California Institute of Technology, Pasadena, CA 91125} 
\altaffiltext{6}{Cornell University,  Astronomy Department, 220 Space Science Building, Ithaca, NY 14853}
\altaffiltext{7}{Lowell Observatory, 1400 Mars Hill Rd., Flagstaff, AZ 86001}
\altaffiltext{8}{University of Texas, Astronomy Department, RLM 15.308, Texas Austin, TX 78712}
 

\section{Introduction}

The atomic and ionic fine-structure lines [\ion{C}{2}] (158 $\mum$) and [\ion{O}{1}]
(63$\mum$) are the dominant cooling lines for neutral interstellar
gas, and ionic fine-structure lines like [\ion{O}{3}]
 (88 $\mum$), (52
$\mum$) and [\ion{N}{2}] (122 $\mum$) are strong coolants in \ion{H}{2}
regions. These lines can be used as diagnostics to infer physical
conditions in the gas, such as temperatures, densities and radiation
fields, by comparing with models of photodissociation regions (PDRs
e.g. Tielens \& Hollenbach 1985, Sternberg \& Dalgarno 1989, Wolfire,
Tielens \& Hollenbach 1990, Kaufman et al. 1999) and \ion{H}{2} regions
(Rubin et al. 1991).

Among these, [\ion{C}{2}] is the most ubiquitous and best studied
line. It was predicted to be the dominant coolant for diffuse neutral
media by Dalgarno \& McCray (1972), but the first detection of this
line was towards the dense star-forming regions M42 and NGC~2024
(Russell et al. 1980). The early observations of external galaxies
were of nearby or IR bright galaxies and showed that [\ion{C}{2}]
emission was associated with dense gas irradiated by ultra-violet (UV)
light from young star-forming regions, often near galactic nuclei
(Crawford et al. 1985, Stacey et al. 1991). Later, observations of
quiescent galaxies like NGC~6946 showed that averaged over the whole
disk, including the HI-rich outer disk, a significant fraction of the
total [\ion{C}{2}] in a galaxy may arise from diffuse ionized or
diffuse atomic gas (Madden et al. 1993, but see Contursi et
al. 2001). Observations of many FIR lines in the Milky Way by FIRAS ,
including [\ion{C}{2}] (158 $\mum$) and [\ion{N}{2}] (205 $\mum$),
indicated that a fair, and possibly dominant, fraction of [\ion{C}{2}]
emission is from diffuse ($n_e=1-5 \ \cc$) ionized gas (Petuchowski \&
Bennett 1993, Heiles 1994). Recent ISO observations have yielded more
surprises. In a previous paper we reported a deficiency of
[\ion{C}{2}] compared with FIR continuum in three normal star-forming
galaxies (Malhotra et al. 1997, Paper I). A similar deficiency was
reported for ultraluminous galaxies by Luhman et al. (1998).

The range of interpretations for the origin and behavior of
the [\ion{C}{2}] line in particular, and FIR cooling lines 
in general, emphasizes the need for studying a suite of FIR
lines in a large sample of galaxies. The Long Wavelength Spectrometer
(LWS, Clegg et al. 1996) on the Infrared Space Observatory (ISO,
Kessler et al. 1996) has made it possible to observe a large number of
atomic and ionic fine-structure lines in the FIR with unprecedented
sensitivity, so that a large sample of galaxies could be
observed.

In this paper we report on and interpret observations of fine
structure atomic and ionic lines observed in 60 normal galaxies. The
sample consists of galaxies whose luminosity is dominated by
star-formation and excludes Active Galactic Nuclei (AGN).  We observed
60 distant galaxies, for which all the FIR emission is within one LWS
beam, as well as 6 nearby, resolved galaxies. Here we report on the
distant sample.

We observed from one to seven atomic and ionic fine structure lines
for each galaxy depending on its FIR brightness. The lines include
[\ion{C}{2}]($158 \mum$), [\ion{O}{1}]($145 \mum$), [\ion{N}{2}]($122
\mum$), [\ion{O}{3}]
($88\mum$), [\ion{O}{1}]($63 \mum$), \ion{N}{3} ($57 \mum$)
and [\ion{O}{3}]
($52 \mum$). The most common lines observed are
[\ion{C}{2}]($158 \mum$), [\ion{O}{1}]($63 \mum$), [\ion{O}{3}]
($88\mum$)
and [\ion{N}{2}]($122 \mum$).  There is no convincing detection of
\ion{N}{3} ($57\mum$), so we do not discuss that line here. This paper
offers a look at the statistical behavior of [\ion{C}{2}],
[\ion{N}{2}], [\ion{O}{3}]
 and [\ion{O}{1}] lines in a diverse sample of
galaxies, aimed at better understanding the physical conditions in the
Interstellar medium (ISM) of normal star-forming galaxies.

The paper is arranged as follows: in Section~2, we discuss the
far-infrared lines observed, their properties, and their diagnostic
value in deriving the physical conditions in the ISM. In Section~3 we
discuss the observations and data reduction procedures. In Section~4
we describe the statistical trends observed; in Section~5 we interpret
these trends, and in Section~6 we discuss the physical conditions in
the PDRs in these galaxies. Section 7 contains conclusions and a
summary of the main results. Appendix A contains a description of the
sample selection and how the sample spans the parameter space in
galaxy properties. Appendices B and C contain tables of line fluxes and
derived physical quantities: far-UV flux $G_0$, gas density $n$, temperature $T$ and pressure $P$.

\section{Far Infrared Fine structure lines}

\begin{table*}[htb]{}
\caption[ ]{Properties of FIR lines observed }
\begin{tabular}{lccccl}
\hline
\noalign{\smallskip}
Species & Wavelength & Excitation  &  Ionization & $T=\Delta E/k $ & $n_{crit}$ \tnm{c} \cr
	&	$(\mum)$ & Potential \tnm{a} (eV) & Potential  \tnm{b} (eV) & (Kelvin) & $\cc$ \cr
\hline
\noalign{\smallskip}
 [\ion{C}{2}]  &  157.714   &  11.26 & 24.38 & 91      & $3 \times 10^3$ [H] \cr
        &            &        &       &         & $5 \times 10^3$ [H$_2$], 50[e]\tnm{1} \cr
 [\ion{O}{1}]   &  145.525 &   -    & 13.62 & 98      & $1 \times 10^5 (T/100)^{-0.57}$ [H] \cr
        &            &        &       &         & $8 \times 10^4 (T/100)^{-0.34}$ [H$_2$] \cr
 [\ion{O}{1}]    &   63.183   &        &       & 228     & $8.5 \times 10^5 (T/100)^{-0.69}$ [H] \tnm{2,3} \cr
        &            &        &       &         & $4 \times 10^5 (T/100)^{-0.34}$ [H$_2$] \tnm{4} \cr
 [\ion{N}{2}]  &  121.89    &  14.53 & 29.6  &         & $3.1 \times 10^2$ [e] \tnm{1} \cr
 [\ion{O}{3}]
 &  88.356    &  35.12 & 54.93 &         & $5.1 \times 10^2$ [e] \tnm{1} \cr
        &  51.815    &        &       &         & $3.6 \times 10^3$ [e] \tnm{1} \cr
 \ion{N}{3} &  57.317    &  29.60 & 47.45 &         & $3 \times 10^3$  [e] \tnm{1} \cr
\hline
\end{tabular}
\tablenotetext{a}{ Potential required to create the ion}
\tablenotetext{b}{ Potential required to ionize the species}
\tablenotetext{c}{ critical density}
\tablenotetext{1}{ Genzel 1991}
\tablenotetext{2}{ Hollenbach \& McKee 1989}
\tablenotetext{3}{ Launay \& Roeff 1977}
\tablenotetext{4}{ Jaquet et al. 1992}
\end{table*}

Table 1 summarizes the properties of the lines observed for this
sample and discussed in this paper (cf. Genzel 1991 for a more
exhaustive description of atomic, ionic and molecular lines in the
infrared). Here critical density for a transition is the density at
which collisions balance spontaneous radiative transitions. $T=\Delta
E/k$ indicates the temperature corresponding to the energy difference
in the upper and lower levels.

\subsection{[\ion{C}{2}]}

The C$^{+}$ fine structure transition at $157.714 \ \mum$ is the most
important coolant of the warm neutral interstellar medium. 
Carbon is the fourth most abundant element and has a lower
ionization potential (11.26 eV) than hydrogen, so that carbon will be
in the form of C$^{+}$ in the neutral surface layers of far-UV
illuminated neutral gas clouds.  The depth of these C$^+$ zones is generally
determined by dust extinction and often extends to $A_V \le
4$. The $158 \ \mum$ [\ion{C}{2}] line is also relatively easy to excite
($\Delta$E/k $\simeq 91$ K), so that $C^{+}$ can cool warm ($30\,K\,<\, T\,<\,10^4\rm\,K$) neutral gas
where the two most abundant atoms H and He cannot (cf. Tielens \&
Hollenbach (1985), hereafter TH85; Wolfire, Tielens \& Hollenbach
1990, hereafter WTH90). In the $C^+$ zones of PDRs, the electrons
are supplied by the $C^+$ ions so that the abundance of electrons
relative to hydrogen is $x(e) \simeq 10^{-4}$. In such conditions the
[\ion{C}{2}](158 $\mum$) line is excited by collisions with hydrogen atoms  

[\ion{C}{2}] from neutral regions can be used in combination with the [\ion{O}{1}]
lines and the FIR continuum, to derive the
gas density $n$ and the incident FUV (6 eV $< \rm h \nu < $ 13.6 eV)
radiation flux G$_0$ respectively (e.g. Wolfire, Tielens \& Hollenbach 1990; Kaufman et al. 1999). $G_0$ conventionally is the 
FUV flux normalized to the average local interstellar flux of $1.6 \times
10^{-3}$ ergs cm$^{-2}$ s$^{-1}$ (Habing 1968).
Since [\ion{C}{2}] comes from both ionized
and neutral regions, decomposing the two components may be necessary
before the comparison with [\ion{O}{1}] can be made.

\subsection{[\ion{N}{2}]}

Since nitrogen has an ionization potential (14.5 eV) higher than that
of hydrogen, [\ion{N}{2}] (122 $\mum$) arises only in ionized gas. This fine structure line is excited by collisions with
electrons. [\ion{N}{2}] (122
$\mum$) has a critical electron density of $3.1\times10^2\,\cc$, and [\ion{C}{2}] (158$\mum$) has a critical electron density of 50 $\cc$ (Table 1). We can estimate
the [\ion{C}{2}] arising from low density ($n_e<50\,\cc$) ionized regions by comparison with the [\ion{N}{2}] (122
$\mum$) line, as described in section 5.4 below.

\subsection{[\ion{O}{1}]}

Oxygen has an ionization potential of 13.62 eV, quite close to that of
hydrogen, so atomic O is found in neutral regions only. In PDRs OI can
exist in atomic form far deeper into clouds than C$^+$. All oxygen not
incorporated into CO can stay atomic to depths as large as A$_V=10$ if
relatively high FUV fluxes impinge on a cloud (TH85). OI has two fine
structure transitions, at $63 \mum$ and $145 \mum$.  The excitation
energy of [\ion{O}{1}](63 $\mum$) corresponds to 228 K and the critical density
is $\simeq 5 \times 10^5 \cc$ (at T $\simeq 300$\ K), so OI lines arise in
warm and dense neutral regions.  As discussed above, [\ion{O}{1}](63 $\mum$),
[\ion{C}{2}](158 $\mum$) and the FIR continuum constrain $G_0$ and $n$. The
$145 \mum$ line lies $\Delta E/k = 325 \ $K above the ground state so
the ratio of [\ion{O}{1}]($145 \mum$)/[\ion{O}{1}] (63 $\mum$) measures temperature
(or $G_0$) in the temperature range of $\simeq 300K$.  In many cases,
however, this ratio may be an indicator of optical depth in the $63
\mum$ line, which is often moderately optically thick. Early
observations and modeling of the [\ion{O}{1}] lines from Orion indicated that
the 63 $\mum$ line was optically thick in emission from the Orion PDR
(Stacey et al. 1983, 1993), while more recent observations of Sgr B2,
DR21 and NGC6334 indicate self absorption in the [\ion{O}{1}] line by
intervening cool O along the line of sight (Keene et al. 1999;
Poglitsch et al 1996; Kraemer et al. 1996).

\subsection{[\ion{O}{3}]
}

Since \ion{O}{2} has an ionization potential of about 35 eV (Table 1), \ion{O}{3}
lines come from ionized regions of galaxies where the UV radiation
field is dominated by fairly early type ($\approx$O6) stars.  The ratio of the
two lines of [\ion{O}{3}]
 at 88 $ \mum$ and 52 $\mum$ is useful for
determining electron density in \ion{H}{2} regions. The $n_e$ derived is
insensitive to the temperature of electrons $T_e$ (Rubin et al. 1994)
and independent of abundance variations. Since the line ratios give a
measure of local density, a large beam measurement by this technique
gives $<n_e>$, as opposed to $<n_e^2>^{1/2}$ provided by emission
measures and observed extent.

\subsection{\ion{N}{3}}

The ratio \ion{N}{3}($57 \mum$)/[\ion{N}{2}](122 $\mum$) provides a measure of
the effective temperature of the ionizing star(s), $T_{eff}$ (Rubin et
al. 1994). \ion{N}{3}($57 \mum$) is only marginally detected in some
sources and we are not sufficiently confident of the measured fluxes
to include \ion{N}{3} in this paper.

\section{Observations and data analysis}

All the line observations for this project were obtained using the
low-resolution grating mode of LWS (Clegg et al.  1996). With the LWS
spectral resolution of 0.29 and $0.6 \mum$ (for wavelength ranges
43-90.5 $\mum$ and 90.5-197 $\mum$, respectively) we do not resolve
the lines.  Sixty distant galaxies were selected to have
$FWHM < 30 \arcsec$ in FIR emission using deconvolved IRAS maps
. Therefore, with the LWS 70\arcsec \ beam, we report the total line
fluxes for these galaxies, and, using measured distances, the total
luminosities from these galaxies \footnote{In practice we will always
use ratios, so the distance errors cancel}. The
observations were made in LWS02 mode where the gratings are scanned
across known lines. A few galaxies were observed in the LW01 mode
where a complete scan was made to cover the wavelength range 45-193
$\mum$.

The line observations were planned to achieve a sensitivity of a fixed
fraction of the FIR continuum flux, depending on the expected line
strength and the feasibility of the observations. Between one and seven of the lines
[\ion{C}{2}](158 $\mum$), [\ion{O}{1}](145 $\mum$), [\ion{N}{2}](122 $\mum$),
[\ion{O}{3}]
(88 $\mum$), [\ion{O}{1}](63 $\mum$), \ion{N}{3} (57 $\mum$) and [\ion{O}{3}]
(52 $\mum$) were observed for each of the 60 galaxies in the sample, depending on
what was practical given the FIR brightness of the galaxy.  The most
commonly observed lines were [\ion{C}{2}]($158 \mum$), [\ion{O}{1}]($63 \mum$),
[\ion{O}{3}]
($88\mum$) and [\ion{N}{2}]($122 \mum$), in that order. Table 2 gives
the planned sensitivities for the different lines along with the
1-$\sigma$ noise actually achieved for these lines averaged over the
whole sample of galaxies.

\begin{table*}[htb]{}
\caption[ ]{Planned and achieved line sensitivities}
\begin{tabular}{lccl}
\hline\noalign{\smallskip}
Line & wavelength & 1-$\sigma$ flux (planned) & Average 1-$\sigma$ flux (achieved)\cr
     &  $\mum$ 	  & \multicolumn{2}{c}{ (as a fraction  of the continuum flux, $\ffir$) } \cr
\noalign{\smallskip}
\hline\noalign{\smallskip}
[\ion{C}{2}]  &    158  &  $2.2\times 10^{-4}$  &   $1.4\times 10^{-4}$ \cr
[\ion{O}{1}]   &    145  &  $2.0\times 10^{-5}$  &   $3.0\times 10^{-5}$ \cr
[\ion{N}{2}]  &    122  &  $8\times 10^{-5}$    &   $7.0\times 10^{-5}$ \cr
[\ion{O}{3}]
 &     88  &  $1\times 10^{-4}$    &   $1.3\times 10^{-4}$   \cr
[\ion{O}{1}]   &     63  &  $1.0\times 10^{-4}$  &   $1.3\times 10^{-4}$	\cr
\ion{N}{3} &     57  &  $1.3\times 10^{-4}$  &   $2.1\times 10^{-4}$	\cr
[\ion{O}{3}]
 &     52  &  $2.4\times 10^{-4}$  &   $2.4\times 10^{-4}$	\cr
\noalign{\smallskip}
\hline
\end{tabular}
\end{table*}

The [\ion{C}{2}] line observations were planned to achieve ($1\sigma$)
sensitivities of $2.2\times 10^{-4}\times \ffir$, where $\ffir$ is the
far-infrared flux of the galaxy between $42 \mum$ and $122 \mum$ and is
computed according to the relation $ \ffir=1.26 \times 10^{-14} [2.58
\times F_{\nu}(60\mum) + F_{\nu}(100\mum)] W m^{-2}$ (Helou et
al. 1988), where $F_{\nu}(60\mum)$ and $F_{\nu}(100\mum)$ are flux
densities in Jansky in the IRAS 60 and 100 $\mum$ bands. For
comparison, previous observations of the Milky Way, starburst galaxies and
galactic nuclei show that the line to continuum luminosity ratio $\rat
= 1-10\times 10^{-3}$ (Stacey et al. 1985, Wright et al. 1991,
Crawford et al. 1985, and Stacey et al. 1991). In Galactic PDRs
associated with \ion{H}{2} regions $\rat$  varies from $\sim 3\times
10^{-3}$ in NGC 2023 (a reflection nebula) to $\sim 8\times 10^{-5}$
in W51, decreasing with \ion{H}{2} region or PDR density and FUV flux (cf. Crawford et
al. 1985, Hollenbach, Takahashi \& Tielens 1991).
 
The data were calibrated with the ISO pipeline OLP7.0 and reduced by
two methods. In the first method the line profiles were derived from
several scans by running a median boxcar filter through them. An
automatic rejection algorithm was used which discarded measurements
more than 5$\sigma$ away from the mean flux level and further
discarded four measurements subsequent to the discarded points. This
sigma-clipping gets rid of cosmic rays and a time-dependent
rejection is done to remove possible memory effects on the
detectors. The median of the observed fluxes is used instead of the
mean to reduce the influence of outlying points arising from cosmic
ray hits. The flux in the lines was determined by directly integrating
under the line, after fitting a linear baseline to the underlying
continuum measurements. The upper limits on non-detections were
derived by calculating the flux from a hypothetical gaussian line with
an amplitude of $3\sigma$ and the effective instrumental profile,
since the lines are unresolved for all sources. In the second method,
the individual scans were inspected and bad points were rejected after
manual inspection.  A linear baseline and a gaussian profile was then
fit to the data.  The line fluxes are derived from the fitted
gaussian. The fluxes from the two methods agreed for well detected ($>
5 \sigma$) sources. The line fluxes thus derived are tabulated in
Appendix B.

Presently, the calibration uncertainties in the flux measurements are
due to two sources: (1) ill-determined dark currents, which are
additive in nature and affect the measured line fluxes minimally
($<5\%$ in most cases) because the dark currents are subtracted with
the continuum levels while determining the line flux; and (2) the
absolute calibration and spectral response is tied to observations and
a model atmosphere of Uranus. The relative flux calibration is
determined by a measurement of a few bright lamps in between
observations. We estimate that the flux uncertainty in our sources is
better than 20\%, including all random errors and systematic
calibration terms. The instrument team reports calibration better that
10-15\% for compact, bright sources observed on-axis (Swinyard et
al. 1999).

\section{Observations and Phenomenology}

\subsection{The [\ion{C}{2}](158 $\mum$) observations} 


While [\ion{C}{2}] and [\ion{O}{1}] are the main cooling lines of the neutral atomic
ISM, the gas heating is dominated by photoelectrons from dust grains
(Watson 1972; for a recent review see Hollenbach \& Tielens 1999). In
a fairly indirect and inefficient mechanism, incident FUV photons with
energies high enough to eject electrons from dust grains (h $\nu > 6
\ eV$) heat the gas via these photoelectrons, with a typical efficiency
of $0.1-1\%$. Efficiency is defined as the energy input to the gas
divided by the total energy of the FUV photons absorbed by dust
grains. This efficiency is determined by the microphysics of the
grains, in particular the work function, photoelectric yield and the
charge of the grains (which is determined by the ratio of FUV fluxes
to gas density G$_0$/n). Before ISO observations little variation was
seen in the total $\rat$ of galaxies (Stacey et al. 1991, Crawford et
al. 1985), or in the Milky Way at 7$\deg$ scales (Wright et
al. 1991).  This suggested that averaged over galaxy scales, the
heating efficiency did not vary much, and guided our line sensitivity
choices, which were scaled to the FIR continuum.

The two circumstances where we do expect the [\ion{C}{2}]/FIR ratio to change
are: (1) for a high ratio of FUV flux to gas density, G$_0$/n, the
grains get positively charged raising the potential barrier for
photoelectric ejection, thereby dropping the heating efficiency; and
(2) when the hardness of the radiation changes, changing the ratio of
FUV light which is effective in photoelectric heating the gas to less
energetic light which can only heat the dust.  In subsequent sections
we will see examples of each of these.

In a previous paper (Paper I) we presented [\ion{C}{2}] measurements for half
the current sample, and saw large variations in $\rat$ including three
non-detections of the [\ion{C}{2}] line in sources with relatively large FIR
continuum fluxes. With a full sample of 60 galaxies and with better
calibrations we verify the main observational results of Paper I. From
our comprehensive data (Table 6 and Figure 1) we observe:

(1) About two-thirds of the observed galaxies (41 out of 60) show a
ratio $\rat > 0.2\%$. This is consistent with models of
photoelectric heating in PDRs illuminated by moderate FUV fluxes
(TH85) and with previous observations (Stacey et al 1991).

(2) There is a trend of decreasing $\rat$ with warmer FIR colors,
$\6f/\f100$ and increasing star-formation activity, indicated by
higher $\lfir/\lb$ ratios, where $\lb$ is the luminosity in the B
band. We test the significance of these correlations by performing the
generalized Kendall's test (Isobe, Feigelson \& Nelson 1986,
Brown, Hollander \& Korwar 1974) which accounts for censoring of data
(i.e. upper limits).  Since we don't know the exact functional form of
the dependence of $\rat$ on $\6f/\f100$ and $\lfir/\lb$, a rank test is
appropriate. That $\rat$ and $\6f/\f100$ are uncorrelated is excluded
at the 99.999\% (4.4 $\sigma$) level. The anti-correlation between
$\rat$ and $\lfir/\lb$ is somewhat weaker and the hypothesis that
$\rat$ and $\lfir/\lb$ are uncorrelated is rejected at the 99.5\% level
($2.8 \sigma$).

(3) Three galaxies (NGC~4418, IC~0860 and CGCG~1510.8+0725) near the
extreme end of this trend in the $\6f/\f100$ ratio and in the ratio of
$\lfir/\lb$ showed no detectable line emission in the [\ion{C}{2}] and [\ion{O}{1}]
lines, as reported in Paper I. Here we report the observation of CGCG~1510.8+0725 with
greater sensitivity and the detection of the [\ion{C}{2}] line, consistent with the
3$\sigma$ upper limit reported earlier. No other prominent lines,
e.g. [\ion{O}{1}] (63 $\mum$), are seen in these galaxies.

\begin{figure}[htb]
\epsscale{1.0}
\plottwo{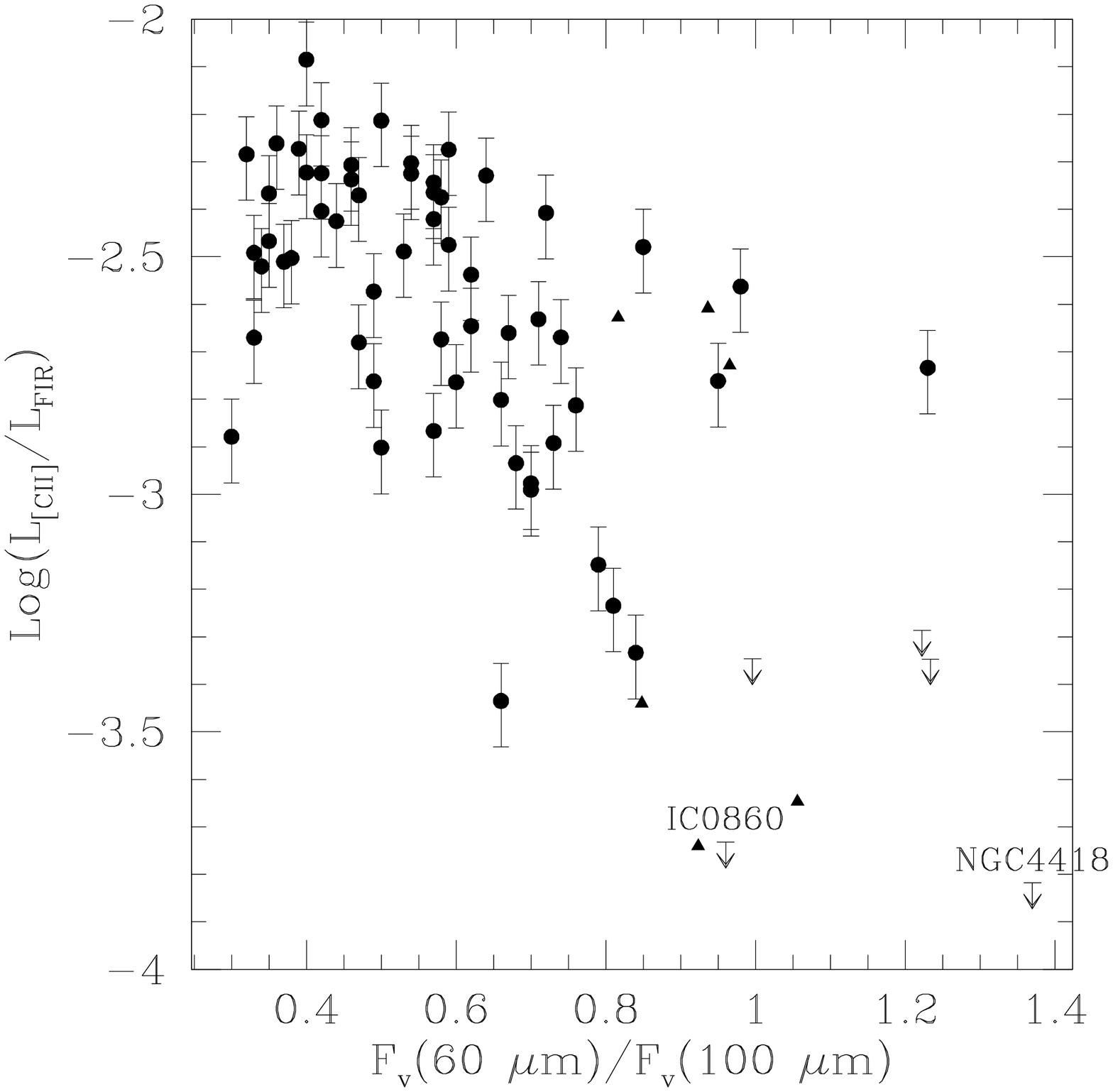}{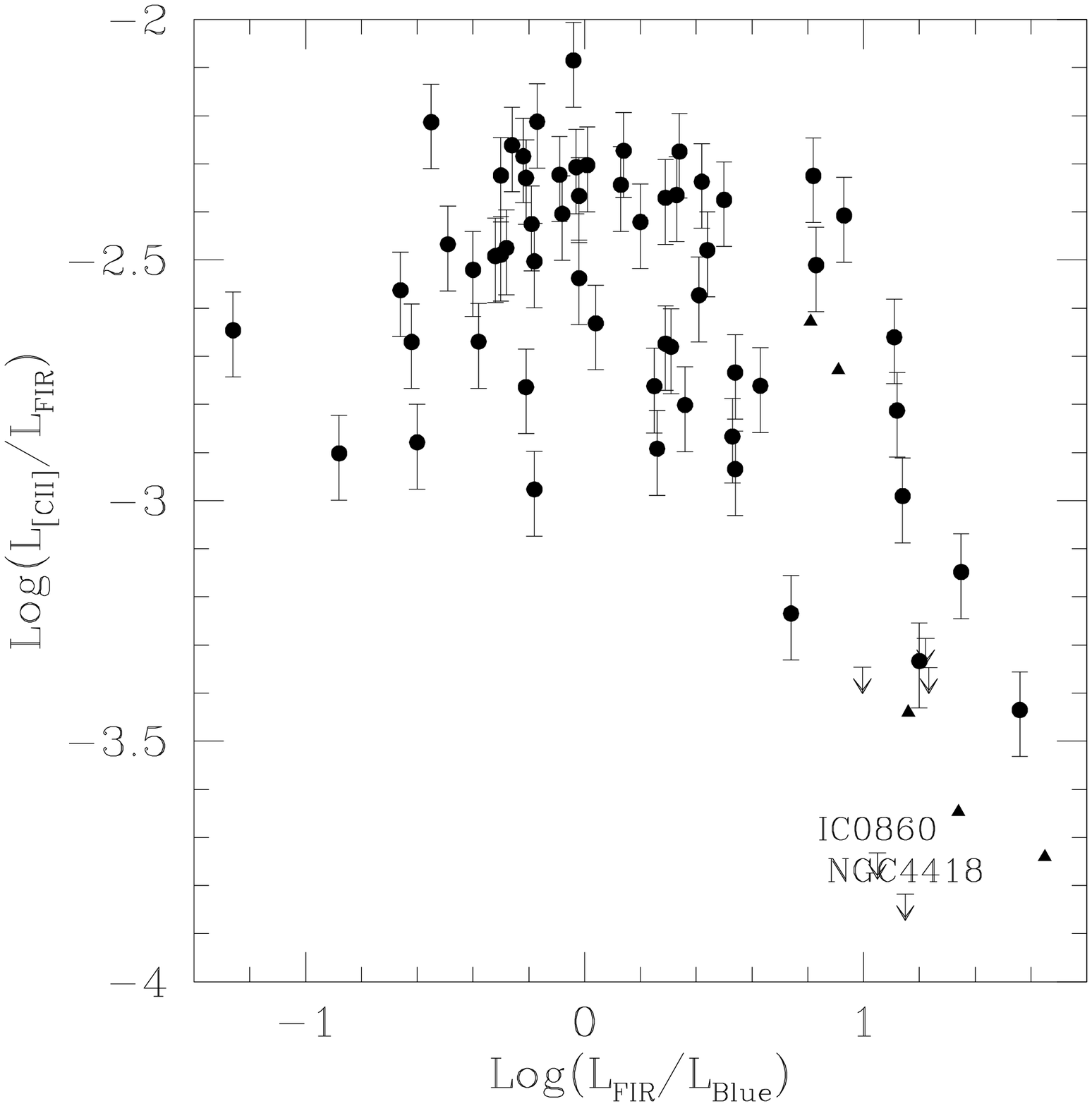}
\caption{(a) The ratio of [\ion{C}{2}] to far-infrared continuum,
$\rat$, is plotted against the ratio of flux in the IRAS $60
\mum$ and $100 \mum$ bands, $\r61$.  Filled circles are normal
galaxies from the ISO-KP sample and triangles are luminous and
ultraluminous galaxies from the sample of Luhman et al. (1998). The line
fluxes are uncertain by about 20\%. There is a trend for galaxies with
higher $\r61$ (indicating warmer dust) to have lower $\rat$, for
normal as well as ULIRGs. Two normal galaxies in a sample of 60 have
no detected [\ion{C}{2}], and they are identified with labels and shown as upper limit
symbols in the figure; other upper limits come from Luhman et
al. (1998). Rank correlation tests show that $\rat$ and $\r61$ are
correlated at the 4.4$\sigma$ level. (b) $\rat$ shows a
similar but weaker trend ($2.8 \sigma$ significant) with the ratio
$\lfir/\lb$, which is an indicator of star formation activity and
optical depth in dust. Galaxies with higher $\lfir/\lb$ (more active
star formation) have lower $\rat$. The decrease in $\rat$ with
$\lfir/\lb$ is not continuous but sets in for galaxies with
$\lfir/\lb>0.8$.}
\end{figure}

Figure 1 also displays data from the sample of luminous and
ultraluminous infrared galaxies (ULIRGs) of Luhman et al. (1998). They
follow the same trends as normal galaxies.

\begin{figure}[bth]
\epsscale{0.5}
\plotone{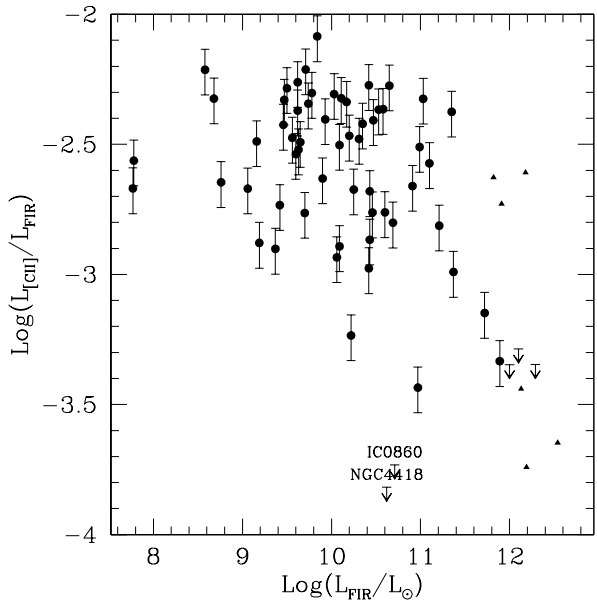}
\caption{ The ratio $\rat$ plotted against the far-infrared luminosity
of galaxies, $\lfir$.  Filled circles represent normal galaxies from the
ISO Key Project sample and triangles denote luminous and ultraluminous galaxies
from the sample of Luhman et al. (1998). There is a trend for galaxies
with higher luminosity to have lower $\rat$. Two normal galaxies in a
sample of 60 and three ULIRGs have no detected [\ion{C}{2}]; they are
identified and shown as upper limit symbols in the figure. The
correlation between luminosity and $\rat$ is weakened by having the
[\ion{C}{2}] deficient normal galaxies in the middle of the luminosity
range.}
\end{figure}

We also explore the dependence of $\rat$ on the total FIR luminosity
of the galaxies (Figure 2). To extend the range of luminosity
explored, we plot the normal galaxies as well as the ULIRGs from the
sample of Luhman et al. (1998). There seems to be a dependence on the
FIR luminosity, but many of the [\ion{C}{2}] deficient galaxies lie in the
middle of the luminosity range and the spread in $\rat$ is large at
the high luminosity end. With only the sample of normal galaxies the
correlation between $\rat$ and luminosity is significant at the $2.1 \sigma$ level;
adding the luminous and ultraluminous galaxies from the sample of
Luhman et al. makes the correlation stronger (3.8 $\sigma$).  We
suspect that the dependence on luminosity is a secondary correlation
and is due to the correlation between FIR colors and luminosity
(Figure 13b in Appendix A).

\subsection{The [\ion{N}{2}] (122 $\mum$) observations}

[\ion{N}{2}] at 122 $\mum$ arises from ionized gas, both in dense \ion{H}{2} regions
and diffuse ionized gas. Figure 3a shows the decrease of $\nf$ with
increasingly warm FIR colors. Unfortunately the detections are fewer
than for the [\ion{C}{2}] (158 $\mum$) and [\ion{O}{1}]($63 \mum$) lines. We detected
this line in about 36\% (19 of 52) of the galaxies observed, and the
entire range of $\nf$, including upper limits, is only about a factor
of 12.  Figure 3a suggests that [NII] follows similar trends as [\ion{C}{2}];
i.e. $\ln/\lfir$ decreases with $\r61$. The anticorrelation between
$\ln/\lfir$ and $\r61$ is $2.5 \sigma$ significant, i.e. there is a 1\%
probability that these two quantities are uncorrelated.
 
The ratio $\cn$ is fairly constant across a
range of FIR colors but with a large scatter: $\cn=9^{+6}_{-6}$
(Figure 3b). A proportionality between the two lines was also found in
spatially resolved observations of the Milky Way (Bennett et
al. 1994). We find an average ratio of $\cn$=9 in sources with [NII]
detections, which agrees very well with the average empirical ratio in
our galaxy (Wright et al. 1991, Bennett et al. 1994). However the many
upper limits on [NII] indicate that many sources have larger $\cn$.

\begin{figure}[bth]
\epsscale{1.0}
\plottwo{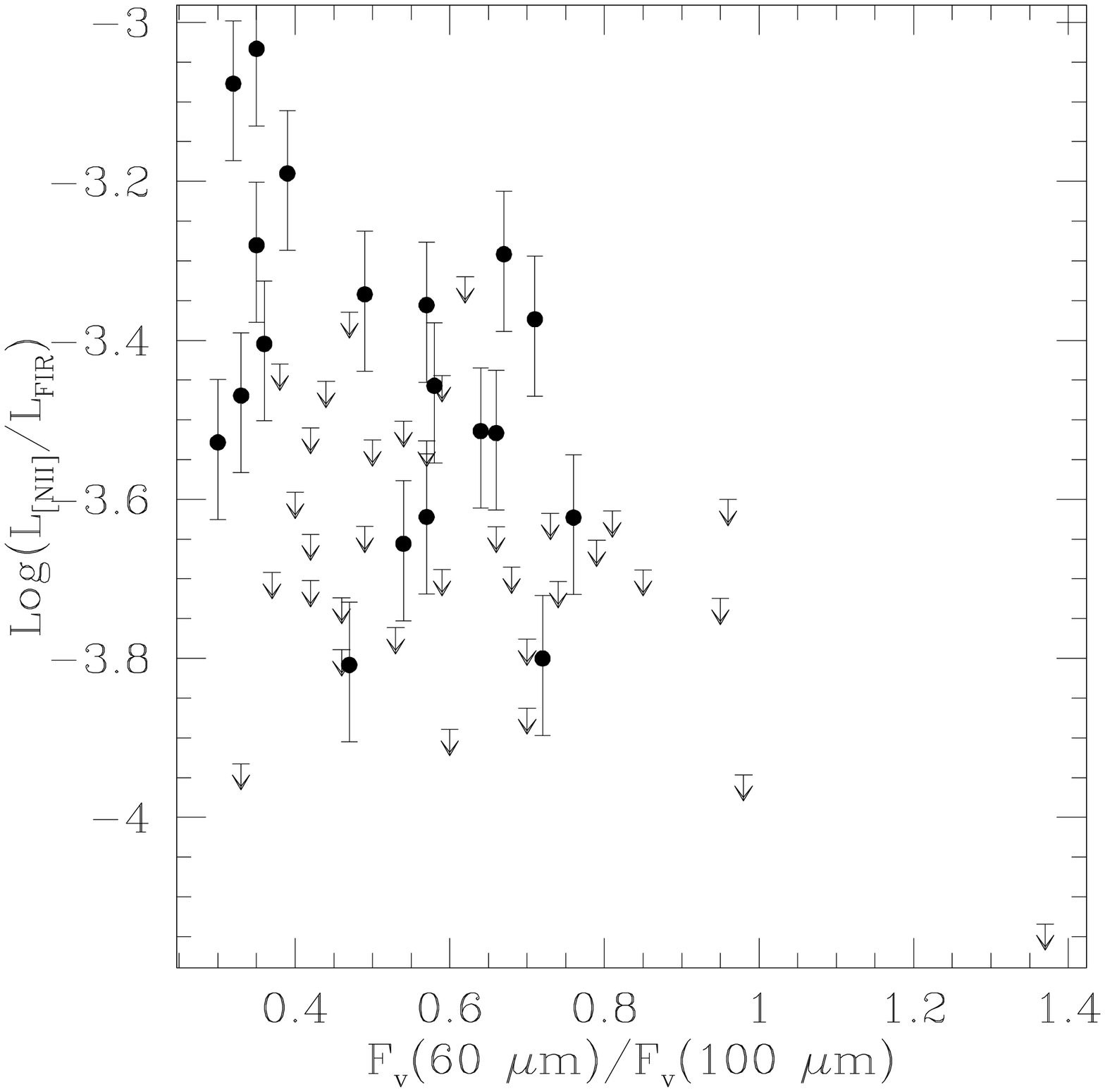}{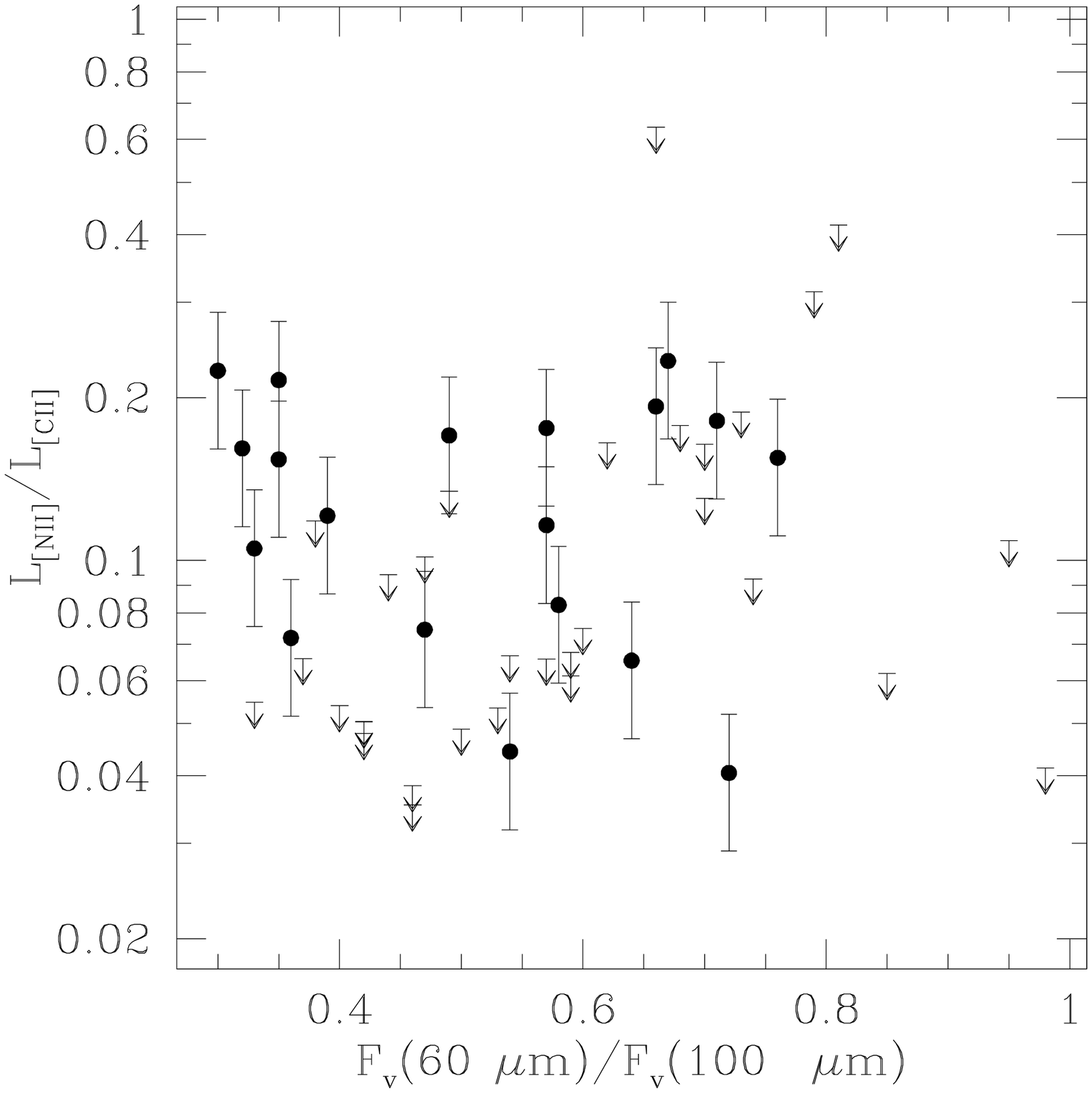} 
\caption{(a) The [NII] (122$\mum$) line shows a fairly similar
behavior to the [\ion{C}{2}] line, i.e. $\nf$ decreases at high $\r61$.  This
anticorrelation is 2.5$\sigma$ significant. If true, this suggests either that
(much of) [\ion{C}{2}] and [NII] arise from \ion{H}{2} regions, or that the PDRs
which produce [\ion{C}{2}] are associated with or surround the ionized gas
which produces [NII]. The upper limits shown in this figure are
$3\sigma$ upper limits. (b) This is further illustrated by the lack of
any observed trends in $\rm L_{[\rm NII]}/\rm L_{[\rm CII]}$ vs
$\6f/\f100$. The observed $\rm L_{[\rm NII]}/\rm L_{[\rm
CII]}=9^{+6}_{-6}$. }
\end{figure}

\subsection{The [\ion{O}{1}] (63 $\mum$ and 145 $\mum$) Observations}

The [\ion{O}{1}] line at 63 $\mum$ is the second most commonly
detected line in the present sample. It is detected in 46 of the 53
galaxies where it was observed. It is not detected in the [\ion{C}{2}]
deficient galaxies. Apart from these galaxies, $\orat$ is unchanging
with $\r61$. The mean value of $\orat = 1.6 \times 10^{-3}$ with a
standard deviation of about 50\%, and the total range of values of
roughly a factor of six (Figure 4). [\ion{O}{1}] (145 $\mum$) is
fainter and was observed/detected only in a few galaxies.

\begin{figure}[htb]
\epsscale{1.0}
\plottwo{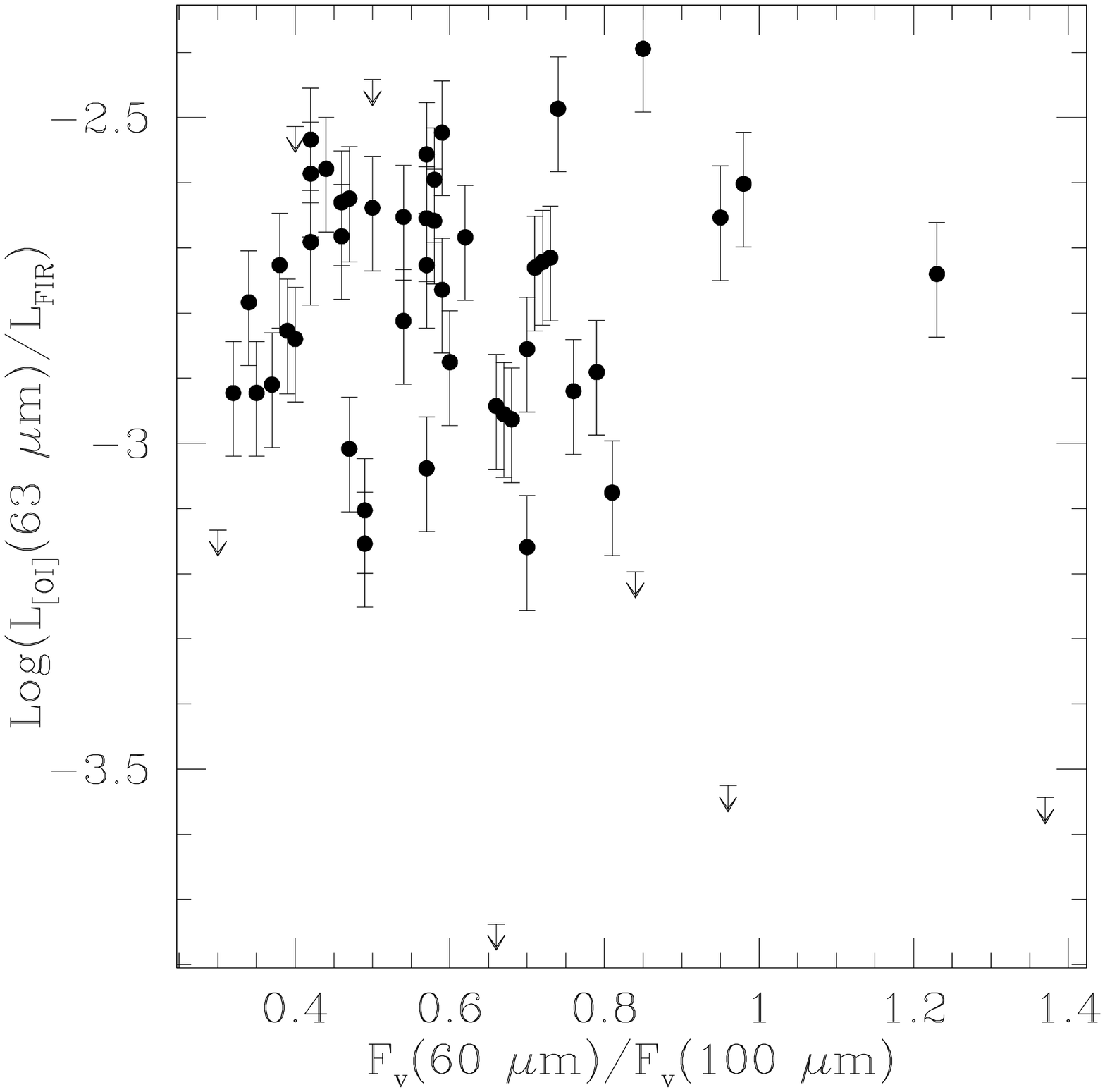}{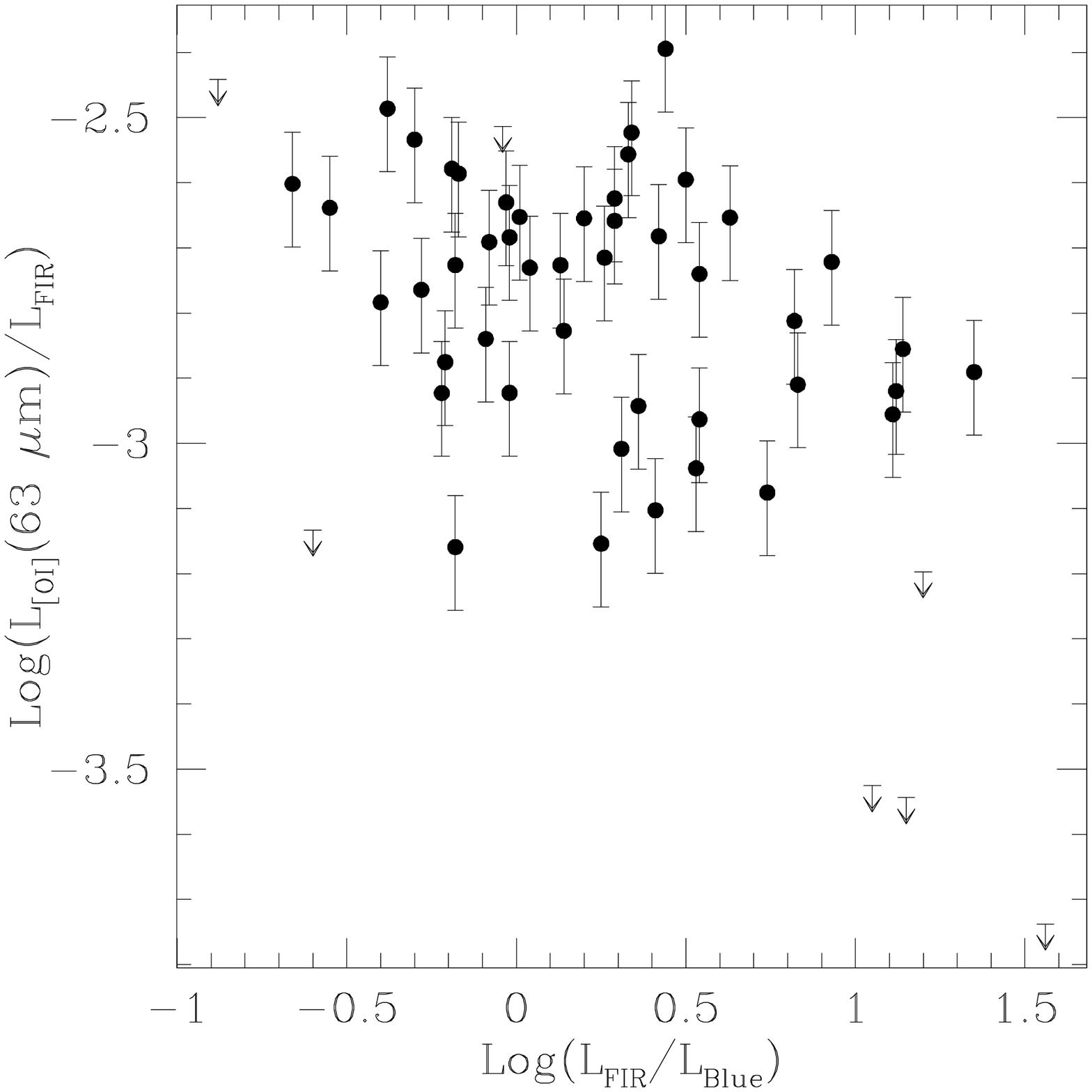}
\caption{$\orat$ shows no trends with FIR colors or FIR-to-blue ratio, 
except for the
non-detections for galaxies with [\ion{C}{2}] deficiency.  In detected sources alone,
the observed $\orat$  ranges over a factor of six, as opposed to a factor of
60 for $\rat$.}
\end{figure}

The [\ion{O}{1}]($63\mum$) line is the other major coolant of neutral regions.
Since the energy to excite this line is higher than for [\ion{C}{2}] we
expect this line to become more important relative to [\ion{C}{2}] in warmer
gas (i.e. higher $G_0$).  This is indeed seen in Figure 5, where
$\ocrat$ increases with $\r61$, indicating that warmer gas correlates
with warm dust. In Figure 1 we saw that $\rat$ decreases with $\r61$
and in Figure 4 we see that $\orat$ does not decrease with $\r61$, so
we should be able to predict the correlation seen in Figure 5. The
generalized Kendall's rank test for the correlation between 
$\lo/\llcii$ and $\r61$ gives a 5.5$\sigma$ significance for a
sample of 50 galaxies. This is the strongest correlation we see in
this dataset, and there is a good physical reason for that.  PDR
models (e.g. Kaufman et al. 1999) show that both the increase in dust
temperatures ($\frat$) and [\ion{O}{1}](63 $\mum$)/[\ion{C}{2}](158$\mum$) 
are due to an increase in FUV flux $G_0$. We make a detailed
comparison between the PDR models and observations in the next
section.

[\ion{O}{1}](63 $\mum$) does not dominate the cooling for the
galaxies in the sample where $\r61\,\le 0.8$. Therefore, if
photoelectric heating is the dominant heating mechanism, $\rat$ tracks
the heating efficiency for the cooler galaxies [$\r61$ $\le
0.8$]. Total heating efficiency is given by the ratio $\rcof$ and
decreases with $\r61$, as discussed in the next section.

\begin{figure}[htb]
\epsscale{0.5}
\plotone{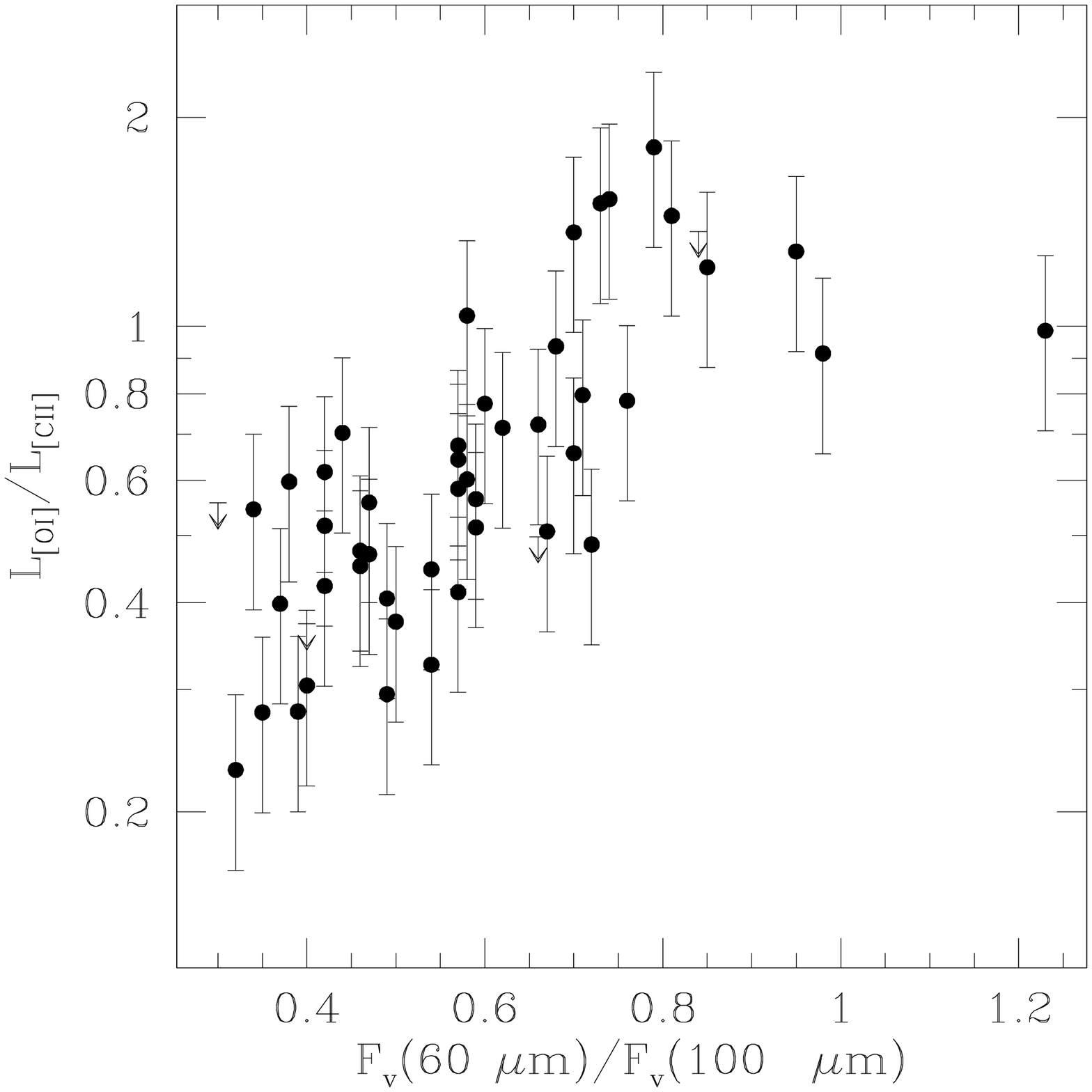}
\caption{The ratio of the main cooling lines for neutral gas,
$\ocrat$, shows a tight correlation with FIR colors (5.5$\sigma$
significant), indicating that warm gas (which emits more [\ion{O}{1}]
(63$\mu$m) relative to [\ion{C}{2}]) correlates with warm dust as indicated by
high $\r61$.  The observed trend is consistent with the PDR models
which ascribe the increase in temperature in gas and dust to higher
FUV flux $G_0$. [\ion{C}{2}] dominates the cooling for galaxies with $\r61$
$<0.8$, and in galaxies with warmer dust [\ion{O}{1}](63$\mum$) becomes more
important.}
\end{figure}

The [\ion{O}{1}](63 $\mum$) line is optically thick in some cases. This can be
seen by comparing it to the [\ion{O}{1}](145 $\mum$) line. The maximum allowed
ratio in the optically thin limit $\rm L_{\rm [OI] 145}/\rm L_{\rm
[OI] 63}$$=0.1$ for $T>300$K (TH85), whereas in the current sample
$\rm L_{\rm [OI] 145}/\rm L_{\rm [OI] 63}$=0.18 for NGC~3620.

The $\rm L_{\rm [OI] 145}/\rm L_{\rm [OI] 63}$ line ratio can also be
used to probe gas temperature; $\rm L_{\rm [OI] 145}/\rm L_{\rm [OI]
63}$ increases with gas temperature, but the S/N in the [\ion{O}{1}]
(145 $\mum$) line is not enough in most galaxies where it is observed.

\subsection{The [\ion{O}{3}]
 ($88 \mum$ and $52 \mum$) lines}

\begin{figure}[htb]
\epsscale{1.0}
\plottwo{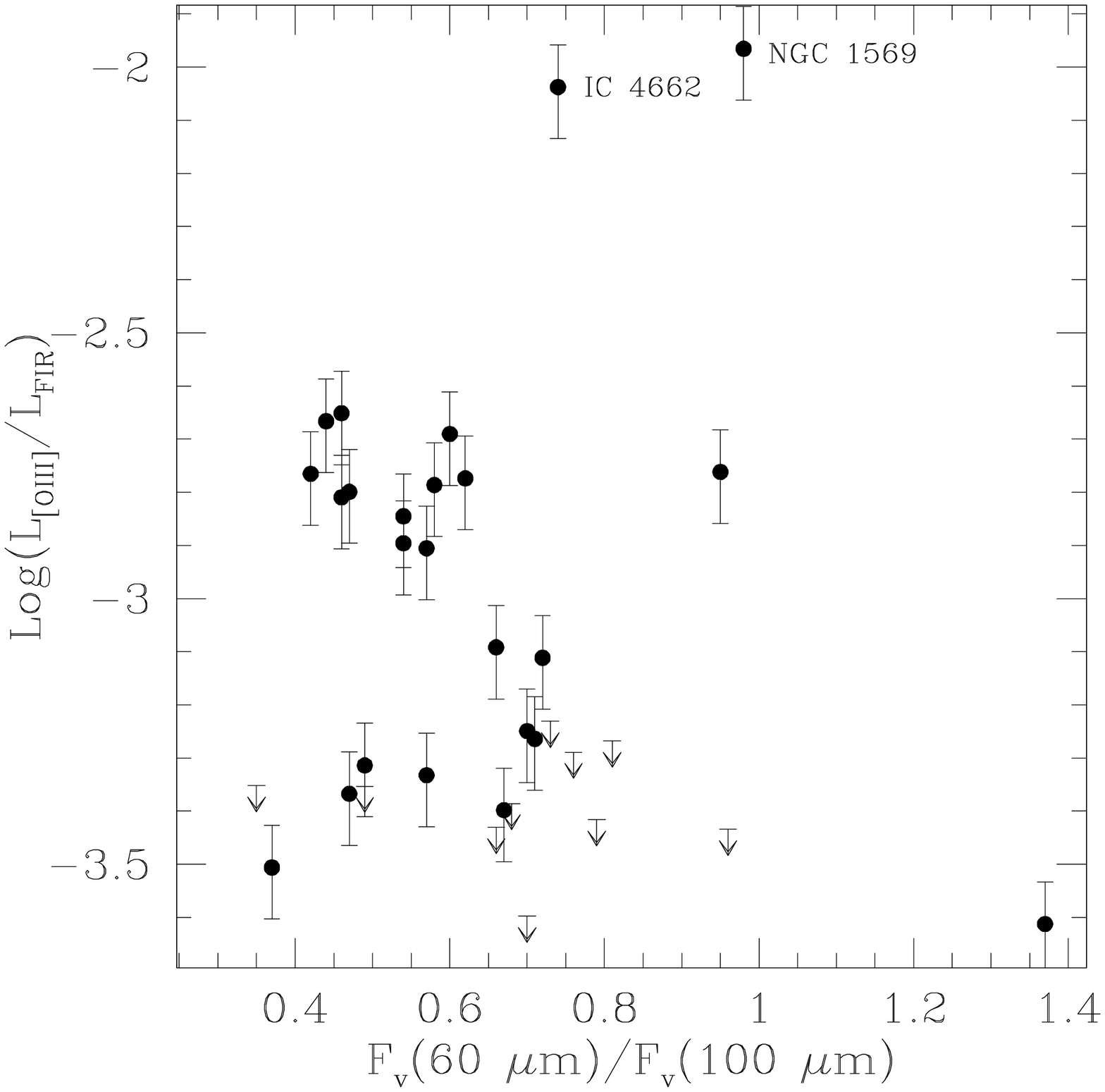}{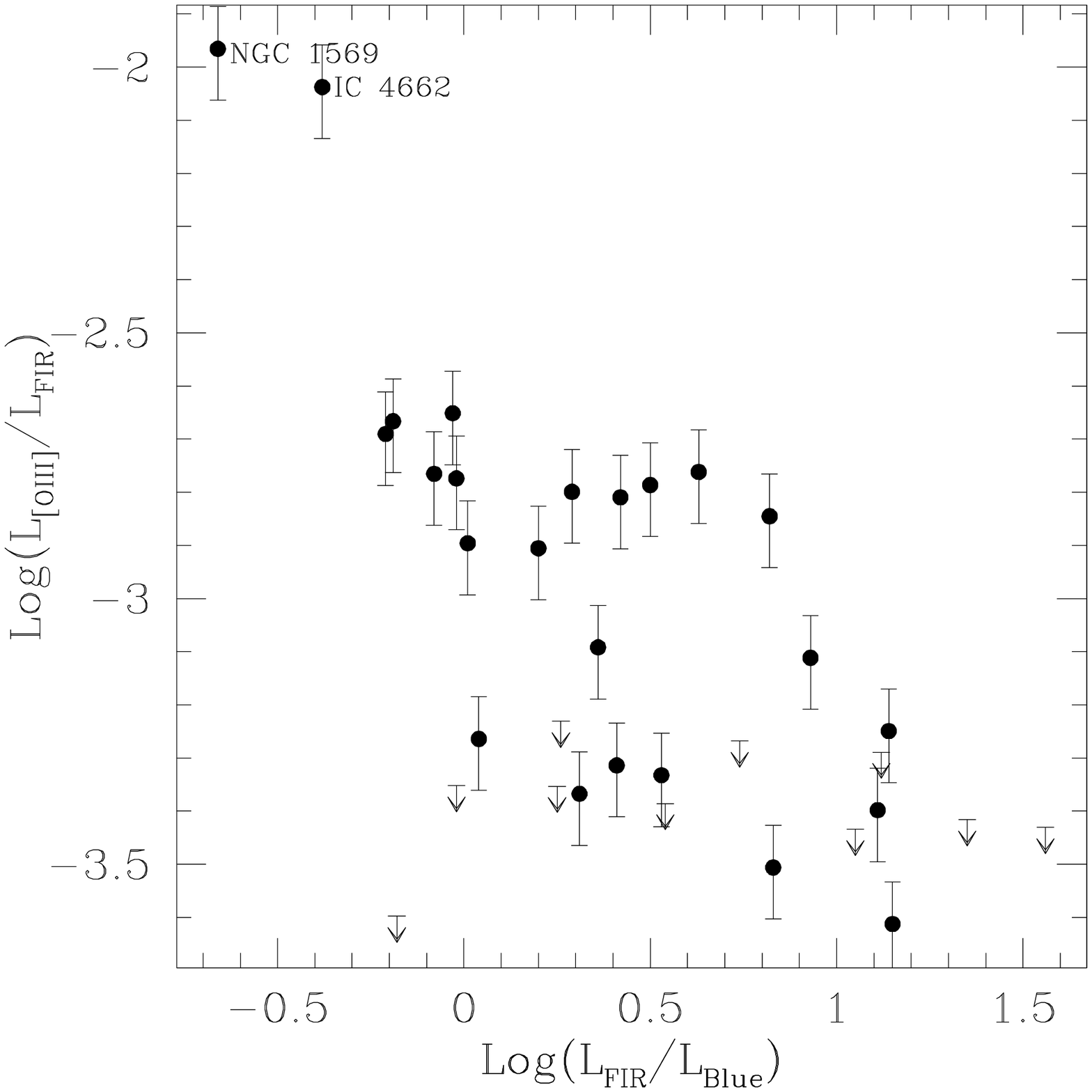}
\caption{$\o3f$ does not vary with FIR colors (the correlation between
$\o3f$ and $\r61$ is only 1.2 $\sigma$ significant). $\o3f$ shows an
anticorrelation with $\brat$ (3.8 $\sigma$ significant), which is
largely contributed by the two irregular galaxies, NGC~1569 and
IC~4662, which have the lowest $\brat$ and have very high $\o3f$. In
fact, [\ion{O}{3}]
 is the strongest far-infrared line in these galaxies.}
\end{figure}

The [\ion{O}{3}]
 line at $88 \mum$ is one of the more easily detected
lines. We detect this line in 24 out of 34 galaxies where it was
observed.  Two low luminosity, irregular galaxies, NGC~1569 and
IC~4662, with low values of $\brat$, have high values of $\o3f$, with
$\rm L_{[OIII]}$ approaching or exceeding 1\% of $\rm L_{FIR}$ (Figure
6). The [\ion{O}{3}]
 (88 $\mum$) line in these galaxies is stronger than the
[\ion{C}{2}] (158 $\mum$) and [\ion{O}{1}] (63 $\mum$) lines from neutral gas. There
is other evidence that the line and continuum emission in these
galaxies may be dominated by \ion{H}{2} regions (see Hunter et al. 2001 for
details). There is relatively little dust near these \ion{H}{2} regions,
hence the low extinction, low $\lfir/\lb$ and high $\o3f$. The
irregular galaxies in the sample which are observed in [\ion{O}{3}]
 have a
high $\o3f$.

The [\ion{O}{3}]
 ($52 \mum$) line is not so widely detected, or observed,
due to the relatively low sensitivity of the detector at those
wavelengths. This line was detected in 3 out of 11 galaxies where it
was observed. 

\section{Interpretation of the data}

\subsection {Electron densities derived from [\ion{O}{3}]
 lines}

For three galaxies, NGC~5713, NGC~1569, and NGC~4490, we use the
measured ratio of the two [\ion{O}{3}]
 lines, at 88 and 52 $\mu$m, to derive
electron densities $n_e$ using the semi-empirical treatment of Rubin
et al. (1994) with an assumed \ion{H}{2} region electron temperature T$_e = 10^4\,$~K. The
results are fairly insensitive to the temperature assumed. The 52 $\mu$m upper
limits in five other galaxies can be combined with the measurement of 88 $\mum$
line flux to yield upper limits on $n_e$ (Table 3).

\begin{table*}[htb]{}
\caption[ ]{$n_e$ from [\ion{O}{3}]
 line ratios}
\begin{tabular}{lcl}
\hline
Galaxy Name & [\ion{O}{3}]
 88/[\ion{O}{3}]
 52 & Log($n_e/\cc$)  \cr
\hline
\noalign{\smallskip}
     NGC~0278  &  $ > 1.75  $  & $< 0.1$ \cr
      NGC~520  &  $ > 0.63 $   & $< 2.6$ \cr
    UGC~02855  &  $ > 0.54 $   & $< 2.7$ \cr
     NGC~1482  &  $ > 1.38 $   & $< 1.8$ \cr
     NGC~1569  &    1.55	& 1.5	 \cr
     NGC~4418  &  $ > 0.21 $ & $<3.4 	$ \cr
     NGC~4490  &    1.25	& 2.0	 \cr
    NGC~5713   &    1.04	& 2.2	 \cr	
\hline
\end{tabular}
\end{table*}

\subsection{Variations in $\rat$ and proposed explanations}

In Paper I we discussed many possible reasons for the extremely low
values of $\rat$ observed in some galaxies. Some of these hypotheses can
be ruled out with the observations of other FIR lines, the most
prominent of these being [\ion{O}{1}](63 $\mum$) and the [NII] (122
$\mum$). We review the various reasons proposed for the [\ion{C}{2}] deficiency in
Paper I and in Luhman et al. (1998) and discuss them in light of current observations.

(A) The [\ion{C}{2}] line could be optically thick in emission. This
hypothesis is supported by the compactness of the mid-IR emission in
the three galaxies with the lowest $\rat$. The [\ion{C}{2}] line becomes
optically thick at $N(\rm C^{+})=5 \times 10^{17} $cm$^{-2}$ for a
velocity width of 4~$\kms$ (Russell et al. 1980). Assuming a velocity
width of 120~$\kms$ for NGC 4418 (Sanders, Scoville \& Soifer 1991)
and $N(C^{+})/N(H) \simeq 1.4 \times 10^{-4}$, optical depth of one in
the [\ion{C}{2}] line is reached for column densities of $N(H)=8 \times 10
^{22}$ cm$^{-2}$, or $A_V=40$ in the PDR gas. The total extinction
towards the nucleus of NGC~4418 is $A_V > 50$.  This would imply an
unusually high ratio of PDR to cold molecular gas.

The biggest uncertainty in estimating the column density of gas needed
to produce optically thick emission is the velocity structure of the
gas; if there is velocity crowding (e.g. near the center or in a bar)
or if most of the self-absorption is local (i.e. in the same cloud or
within a small velocity interval), even a small column density of gas
can be effective. But large optical depths are difficult to produce
because of the following argument. Models of PDRs, assuming that grain
opacity and gas abundances both scale with metallicity, show that the
column of C$^+$ in a PDR is only of order a few times $10^{17}$ so
that the [\ion{C}{2}] has optical depths in a given cloud only of order
unity, unless every line of sight through the cloud goes through many
PDRs. But in that case one would see emission from the first optical
depth of unity, so for an optical depth of 100 in C$^+$, one would get
1/100th of the emission compared to the optically thin case (and not
$e^{-100}$). So to explain $\rat$ which is 100 times lower, one would
need an optical depth of 100 in the PDRs, which corresponds to a
hydrogen column density of $\simeq 3 \times 10^{23}\,\rm cm^{-2}$.

Another way to produce large optical depth is to have cold C$^+$ on
the outside of the cloud while the PDR inside produces [\ion{C}{2}]
emission. The cold gas must then be turbulent enough to have a
velocity width equal to or larger than the warm/emitting gas so that
[\ion{C}{2}] emission in the line wings does not escape. Self-absorption is
seen in the $158 \mum$ line when observed at high spectral resolution
(Boreiko \& Betz 1997), and in ISO observations of Sgr~B2 (Cox et
al. 1999), but the line width of the cold gas is usually smaller than
that of the warm gas, leading to a reduction in flux by a factor of
roughly two, but cannot produce a factor of 60 reduction in flux without a
good match in velocity widths.

Since [\ion{O}{1}] is present up to higher optical depths in clouds than
C$^+$, [\ion{O}{1}] is more likely to have larger columns and to be self
absorbed. Towards the Galactic center a deeper absorption profile is
seen in the [\ion{O}{1}] (63 $\mum$) line (Keene et al. 1999, Cox et
al. 1999, Poglitsch et al. 1996, Kraemer, Jackson \& Lane, 1996) than
in the [\ion{C}{2}] line (Cox et al. 1999).  [\ion{O}{1}] should go optically thick
{\it faster} than [\ion{C}{2}] if optical depth were the reason for
decreasing $\rat$. Thus we should have seen a more dramatic decline in
$\orat$ than is seen in $\rat$ with increasing $\frat$ or $\lfir/\lb$.
Since no decline is seen in $\orat$ (Figure 4b), we discount
self-absorbtion as the dominant reason for the decrease in $\rat$.

(B) High dust extinction: For dust
extinction to be important $\tau_{dust}=1$ at 158 $\mum$ implies
$A_V=15,000$. This is still insufficient to explain the [\ion{C}{2}]
deficiency and is unlikely even for very obscured starbursts like
Arp~220 (Fischer et al. 1999). Since the [\ion{O}{1}] line at $63 \mum$ is at
 shorter wavelengths it is even more vulnerable to extinction effects
but still requires far too high an $A_V$ for this to be a viable
explanation. We should have also seen a dramatic decrease in $\orat$
with increasing $\brat$ in figure 4b since $\brat$ is a measure of extinction
effects in galaxies, but such a trend is not seen.

(C) Softer radiation from older/less massive stellar populations could
lead to lower $\rat$ due to a lack of C-ionizing photons. If much of
the grain heating is by longer wavelength photons, rather than by FUV,
[\ion{C}{2}] will be relatively weak compared to FIR because grain
photoelectric heating of the gas will be inefficient. This can happen
when dust heating is by old, low mass stars. Luhman et al. (1998)
postulate this scenario for [\ion{C}{2}]-deficient ULIRGs. An aging
starburst and a weak UV field relative to a softer, optical/infrared
field which heats the dust, is also used to explain the [\ion{C}{2}] line weakness
in the Galactic center (Nakagawa et al. 1995).  This mechanism seems
unlikely to explain the most dramatic [\ion{C}{2}] deficiency systematically
found in galaxies with warmer FIR colors and higher $\lfir/\lb$
ratios, with the hotter dust pointing to the presence of massive stars
and UV photons. It also seems unlikely that all starbursts are old by
the time we see them. This explanation is more likely for the decrease
seen in $\rat$ in early type galaxies with low rates of star formation
and the lowest $\lfir/\lb$ in the sample (to the left extreme of
Figure 1b; also see Malhotra et al. 2000). Softer radiation fields in early
type galaxies then lead to somewhat lower (by factors of a few) $\rat$;
this is corroborated by lower values of [\ion{C}{2}]/CO seen in one of these 
galaxies (Malhotra et al. 2000).

(D) An obscured AGN can lead to both higher $\brat$ and lower
$\rat$. This hypothesis is supported by the observed compactness of
the [\ion{C}{2}]-deficient sources in the mid-IR (Dale et al. 2000). A
buried AGN could produce a small $\rat$ since the UV field is
inefficient at making [\ion{C}{2}], both due to its hardness (higher
ionization states for C will be common) and due to its overall
strength (see (F) below). This explanation may not hold for the
ULIRGs, e.g. Arp 220, which are now believed to be powered by
starbursts (Genzel et al. 1998).  The AGN hypothesis may explain the
deficiency of [\ion{C}{2}] but does not easily explain the trends in $\rat$
with $\lfir/\lb$ and $\r61$ unless a significant fraction of galaxies
have obscured active nuclei, and even then one would expect the AGNs
to add more scatter than a trend.

(E) For high values of $G_0$ (or $T$) and $n$, [\ion{O}{1}] (63$\mum$) emission
is stronger than [\ion{C}{2}] emission, since [\ion{C}{2}] is quenched by
collisional de-excitation. The grain charge, which determines the gas
heating efficiency as measured by the observed ratio $\rcof$, is set
by $G_0/n$. Therefore, with $G_0/n$ fixed, raising $G_0$ and $n$
together does not change $\rcof$, but it does increase $\ocrat$ and
$\orat$ while decreasing $\rat$. In other words, the heating
efficiency stays constant, but the cooling increasingly emerges in the
[\ion{O}{1}] line instead of the [\ion{C}{2}] line as $G_0$ rises with $G_0/n$
fixed. In section 4.3, we observed that the ratio of the two major
cooling lines of the neutral ISM, $\ocrat$, increases with increasing
$\r61$ (Figure 5). $\r61$ is a rough indicator of the dust
temperature and increases with the FUV flux $G_0$. Since the [\ion{O}{1}]
(63$\mum$) line has an excitation energy $\Delta$E/$k$=228 K compared with
$\Delta$E/$k$=92 K for the [\ion{C}{2}] (158$\mum$) line, an increase in gas
temperature caused by an increase in $G_0$ will increase
$\ocrat$. Therefore, the correlation between $\ocrat$ and $\r61$ is
likely due to an increase in $G_0$, which increases both gas and dust
temperatures. However, Figure 7 shows that $\rcof$ does {\it not} stay
constant with increasing $\r61$ or increasing $G_0$. This is a clear
indication that the density is not rising as rapidly as $G_0$ and that
grains are becoming more positively charged. Therefore this
hypothesis, although it may contribute somewhat to the [\ion{C}{2}]
deficiency, cannot be the dominant cause.

(F) For high ratios of FUV flux to gas density ($G_0/n$), the dust
grains become highly positively charged and are less efficient at
heating the gas because of a higher potential barrier to photoelectric
ejection (TH85).  Hence there is less total gas cooling relative to
the FIR continuum emitted by grains. For example, increasing $G_0$
while keeping $n$ fixed raises the grain charge, lowers the gas
heating efficiency $\rcof$, lowers $\rat$ and raises $\ocrat$. An
increase in $G_0$, accompanied by an increase in $G_0/n$, is the
likely explanation for the trend of decreasing $\rcof$ with $\r61$
seen in Figure 7, as the dust gets warmer with increasing $G_0$
and more positively charged with increasing $G_0/n$. In addition, it
accounts for the decrease in ${\rm L_{\rm [CII]}}/{\rm L_{\rm TIR}}$
with $\r61$ seen in the data
plotted in Figure 9. This, then, is our favored hypothesis; in the
next section we compare PDR models to $\rat$ to further bolster this
hypothesis.

\begin{figure}[!htb]
\epsscale{0.5}
\plotone{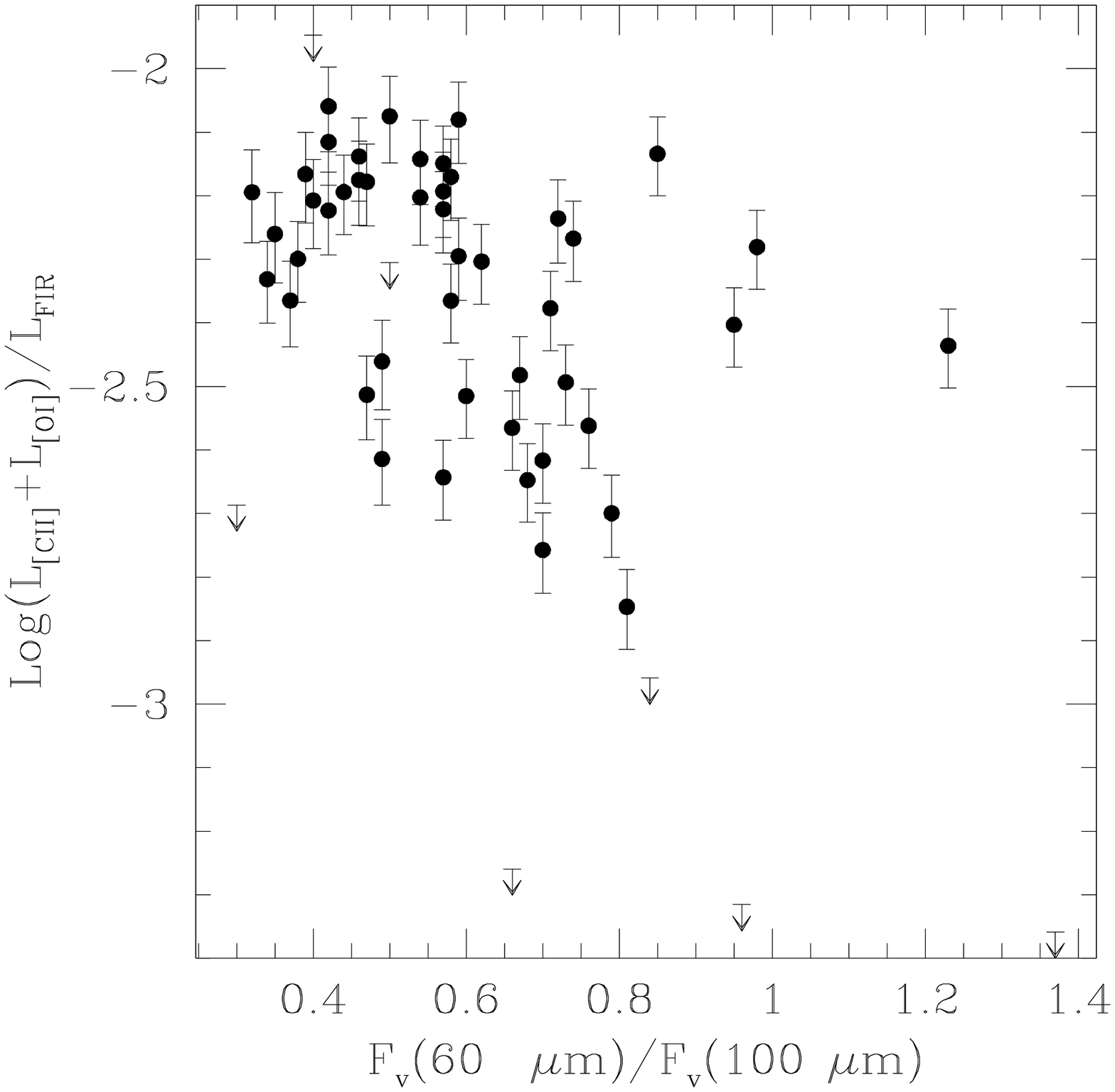}
\caption{Heating efficiency $\rcof$ vs grain temperature $\frat$. A
decline in the total cooling in these lines with respect to FIR is
seen in galaxies with warmer dust temperatures as indicated by
$\frat$.  This is expected as grains become more positively charged in
regimes of high $G_0/n$. The rank correlation test yields that the
anticorrelation between $\rcof$ and $\frat$ is 3.3$\sigma$
significant.}
\end{figure}

\subsubsection{ [\ion{C}{2}] deficiency Case Study: Arp~220}

Of the explanations suggested above, (C) \& (F) postulate a decrease
in neutral gas heating relative to grain heating. In the others, either [\ion{C}{2}] is attenuated by dust (B) or the cooling emerges in other lines (A, D and E). We have already ruled out E (cooling by [\ion{O}{1}]) as the dominant cause of the reduction in $\rat$. We 
can further test hypotheses A and D by measuring the emission in all the major cooling lines to
see which, if any, line has replaced [\ion{C}{2}] and [\ion{O}{1}]  as the coolant of the
neutral medium. The total cooling budget can then be computed to see
if the heating efficiency has declined. (The test is less stringent for D, since the gas cooling may be predominantly in optical lines which are absorbed by dust and emerge as continuum.)

At present Arp~220 is the only [\ion{C}{2}] deficient galaxy for which a
relatively complete census of gas cooling lines has been made. Gerin
and Phillips (1998) have observed this galaxy in the [CI] (492 GHz) line and
higher-lying CO rotational lines, and Sturm et al. (1996) observed this
galaxy in $\rm H_2$ rotational lines. Table 4 summarizes the energy output
in cooling lines of neutral gas in Arp~220.

\begin{table*}[htb]{}
\caption[]{Arp 220: Energy Budget}
\begin{tabular}{ccc}
\hline\noalign{\smallskip}
& Flux (observed) ($W/m^{-2}$)& Flux(extinction corrected) ($W/m^{-2}$)\cr
\hline\noalign{\smallskip}
        FIR Continuum \tnm{1}  &       $6.8 \times 10^{-12} $ &\cr	
        [\ion{C}{2}]        \tnm{1}     &       $8.7 \times 10^{-16}  $&\cr
        CI ($^3P_1$-$^3P_0$) \tnm{1} &  $1.9 \times 10^{-17} $ &\cr
        CI      (total) \tnm{1} &      $1.0 \times 10^{-16} $ &\cr
        CO (1-0)   \tnm{1}     &       $1.8 \times 10^{-18} $ &\cr
        CO (total) \tnm{1}      &      $2.1 \times 10^{-16} $ &\cr
        $H_2$ S(5) \tnm{2}     &$2.4 \times 10^{-16} $ &$2.4 \times 10^{-15}  $\cr
        $H_2$ S(2) \tnm{2}       &$<1.5 \times 10^{-16} $        &$<7.0 \times 10^{-16}  $\cr
        $H_2$ S(1) \tnm{2}       &$2.3 \times 10^{-16} $ &$9.7 \times 10^{-16}  $\cr
        $H_2$ S(0) \tnm{2}       &$<3.5 \times 10^{-16} $        &$<7.3 \times 10^{-16} $\cr
\noalign{\smallskip}
\hline
\end{tabular}
\tablenotetext{1}{ Gerin \& Phillips 1998}
\tablenotetext{2}{ Sturm et al. 1996}
\end{table*}

We see that only the molecular hydrogen lines come close to matching even
the reduced output in [\ion{C}{2}]. [\ion{O}{1}] ($63 \mum$) 
is seen in absorption (Fischer et al. 1999).  We conclude that there is a reduced ratio of gas heating to
grain heating in at least this [\ion{C}{2}] deficient galaxy. This is then consistent with hypotheses C and F. 

\subsection{ Trends in $\rat$: comparison to the PDR models}

Of the various explanations for the trends and the deficiency of
$\rat$ listed above, the data most favor scenario (F) where the
decrease in $\rat$ is due to reduced heating via charged grains at
higher values of $G_0/n$, with scenario (E) contributing to the
decrease in [\ion{C}{2}] luminosity when [\ion{O}{1}](63 $\mum$) starts dominating
($\r61$ $\ge 0.8$). In Figure 9a and 9b we see the quantitative
comparison of $\rat$ and $\orat$ with the PDR models of Kaufman et
al. (1999).  The term PDR refers to the neutral ISM in galaxies,
including diffuse clouds or cold neutral medium (CNM), warm neutral
medium (WNM), dense neutral gas around \ion{H}{2} regions and molecular clouds
illuminated by the interstellar radiation field (ISRF). PDRs are formed wherever FUV photons play a
dominant role in the heating and chemistry of the gas.

In order to compare the data to the models we need to correct the
quantity $\rm L_{\rm FIR}$ used previously to include the dust
emission outside the wavelength range 40-120$\mum$. Since the long
wavelength part of the dust emission ($\lambda > 120 \mum$) is not
measured for these galaxies, we have to rely on dust emission models
based on Galactic dust and other well-studied galaxies. Empirical data
from IRAS and ISO indicate that the overall spectral shape in the
infrared may be represented to first order as a function of one
parameter, the $\frat$ ratio.  Dale et al. (2001) have proposed a
single parameter family of spectral energy distributions (SEDs) based
on those data from which we convert FIR (dust emission in the
40-120$\mum$) to total infrared emission (TIR; 3-1100 $\mum$),
including emission from PAHs and dust associated with cold cirrus. The
uncertainty in converting FIR to TIR is about 15\% (D. Dale, private
communication).

\begin{figure}[!htb]
\epsscale{0.5}
\plotone{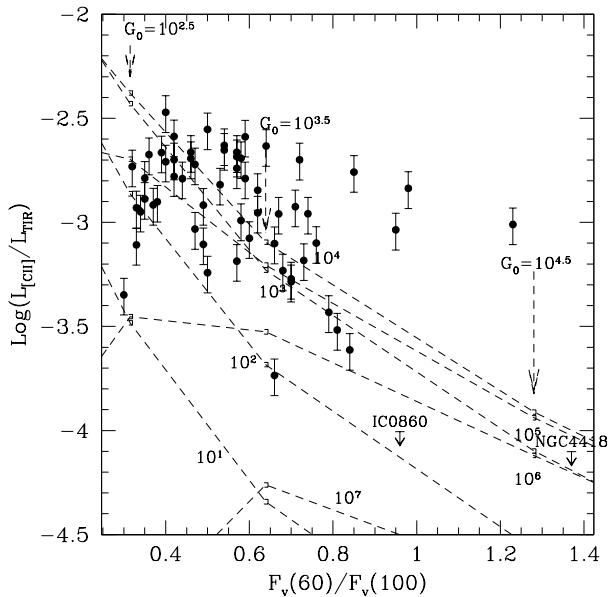}
\caption{ Comparison of $\2rat$ with PDR models. TIR is the total IR
emission (3-1100 $\mum$; see text). The solid dots represent the
$\2rat$ values for the 60 galaxies in this sample. The curves
represent $\2rat$ values for gas densities from $10^1$ to $10^7\,\rm cm^{-3}$, with notches
indicated for the progression of $G_0$, from the PDR models of Kaufman et
al. 1999. Lines of constant G$_0$ are vertical in this plot. The FIR
colors become warmer for higher $G_0$. We see that the trend of
decreasing $\2rat$ with $\r61$ is explained by the models as
being due to the increase in $G_0$.}
\end{figure}

Two things are apparent from this comparison. First, the PDR models do
reproduce the decrease in $\2rat$ with $\r61$. For the extremely
deficient $\2rat$ galaxies, the PDR models of Kaufman et al. (1999)
require $G_0/n\,>10^2\, cm^3$ in order for $\2rat < 10^{-4}$. We note
in Figure 8 that a second PDR solution for $\2rat < 10^{-4}$ occurs
for a much higher $G_0/n$ ratio. However, this solution produces far
more $\lo$ than is observed; it is essentially hypothesis (E) ruled
out from the last section. In the regime $G_0/n\,>10^2\, cm^3$,
radiation pressure on grains may drive them through the gas at large
drift velocities, resulting in additional gas heating, a detail not
considered in the Kaufman et al. (1999) models; recent calculations by
Weingartner \& Draine (1999) show that such grain drift will only be
important for large grains, while small grains, which are responsible
for gas heating and much of the FUV absorption in the PDR models, will
not be significantly affected.

Second, we note in Figure 8 that about half the galaxies have
$\2rat$ which is too high by a factor of about two to be explained by
PDR models with any parameters.  This is not very surprising, since
not all of the [\ion{C}{2}] comes from PDRs; ionized gas in diffuse ionized
regions can contribute significantly to [\ion{C}{2}] emission as well (Petuchowski \&
Bennett 1993, Heiles 1994). We discuss this further in the next
section.

\subsection{[\ion{C}{2}]/[NII](122 $\mum$) from ionized regions}

The ionized gas contribution to [\ion{C}{2}] can be estimated from
measurements of the [NII] (122 $\mum$) line, which arises exclusively
in the ionized gas. For the asymptotic high electron density limit
($n_e >> n_{crit}$, where $n_{crit}= 3.1\times 10^2 \cc$ for
[NII](122 $\mum$) and $n_{crit}=50\,\cc$ for [\ion{C}{2}](158$\mum$)) (see
Table 1):

$$\frac{\rm L_{\rm [CII]}}{\rm L_{\rm [NII]}}= 0.288 \frac{N(C^+)}{N(N^+)}$$
where $N(N^+)$ and $N(C^+)$ are the numbers of the respective ions, and the
Einstein A coefficients used are $A_{158}=2.3 \times 10^{-6} s^{-1}$
(Froese-Fischer \& Saha 1985) and $A_{122}=7.4 \times 10^{-6} s^{-1}$
(Froese-Fischer 1983).

In the low density limit ($n_e << n_{crit}$) the ratio is given by:

$$\frac{\rm L_{\rm [CII]}}{\rm L_{\rm [NII]}}=3.05 \frac{N(C^+)}{N(N^+)}$$
where the collision strengths are taken to be $\Omega_{12}=2.15$
(Blum and Pradhan 1992) and $\Omega_{13}=0.272$ (Lennon and Burke
1994).

The ratio of the [\ion{C}{2}](158$\mum$) and [NII]($122 \mum$)
luminosities depends on the gas phase abundances of these ions. We
expect the $C^+$ and $N^+$ abundances to be different in different
galaxies in the sample. At present, the abundances in these galaxies
are not known so we use Galactic (Solar, or interstellar) abundances.
Even in the Galaxy the differences are substantial: $[C/H]= 1.4 \times
10^{-4}$ in absorption line measurements of diffuse gas (Sofia et
al. 1997) and $[C/H]= 2.4 \times 10^{-4}$ (Esteban et al. 1998, Rubin
et al. 1993) as measured in emission in Orion. Nitrogen abundances are
less discrepant: $[N/H]=7.5 \times 10^{-5}$ from absorption line
measurements of the diffuse gas (Meyer et al. 1997) and $[N/H]=6.8
\times 10^{-5}$ from emission line studies in Orion (Esteban et
al. 1998, Rubin et al. 1993).  For various combinations of those
abundances, the expected $\rm L_{\rm [CII]}/\rm L_{\rm [NII]}$ ranges
from 5.7 to 10.7 in the diffuse ($n << n_{crit}$) limit and from 0.54
to 1.0 for dense ionized gas ($n >> n_{crit}$).  Furthermore,
combining the abundances observed in the diffuse ionized gas (DIG)
with the diffuse gas estimates points to 5.7 as the most reasonable
value to adopt.


In the galaxies where both lines are detected, $\rm L_{\rm [CII]}/\rm
L_{\rm [NII]}$ ranges from 4.3-24, and including upper limits
increases the range to 4.3-29. The geometric mean of $\rm L_{\rm
[CII]}/\rm L_{\rm NII}$ ratio in the sub-sample where both lines are
detected is 8, and the arithmetic mean in 9.

The total [\ion{C}{2}] in a galaxy comes from PDRs, DIG
and \ion{H}{2} regions, whereas the [NII] arises from DIG and \ion{H}{2} regions.
Thus the variable fractions of the galaxy's ISM involved in each of
those components will add scatter to $\rm L_{\rm [CII]}/\rm L_{\rm
[NII]}$, as will variations in C and N abundances.  Using a constant
multiplied by the [NII] flux to correct the [\ion{C}{2}] flux for ionized
region contributions is therefore an oversimplification we adopt in
the absence of further information on individual galaxies.  The
observed values of $\rm L_{\rm [CII]}/\rm L_{\rm [NII]}$ cluster
clearly near the low end, with more than half between 4.5 and 6,
suggesting that the most generally applicable correction to [\ion{C}{2}] must
be near the lower end of that histogram as well.  This also favors the
ratio of 5.7 based on the lower abundance of $C^+$, which we adopt in
what follows.

The amount of [NII] emission from diffuse versus dense regions has
only been measured for the Milky Way. From FIRAS measurements of the
two [NII] lines at 205 and 122$\mum$ in the Milky Way, Bennett et
al. (1994) conclude that most of the emission is from diffuse gas. The
predicted ratio of these lines is I(122)/I(205)=0.7 for diffuse gas
($n_e << 100 \,\cc$), and I(122)/I(205)=3 for $n_e \simeq 100 \cc$
(Rubin 1985). Taking the middle of the range of the observed ratio
I(122)/I(205)=1.0 to 1.6 (Wright et al. 1991), about 75\% of [NII]
emission comes from diffuse gas. This implies that [\ion{C}{2}] emission from
the ionized medium $[CII]_{ion}= 4.3 \times [NII](122)$. We will
therefore approximate the [\ion{C}{2}] flux from PDRs as $[CII]_{c}=
[CII]-4.3\times [NII](122)$ in what follows.

The mean and median adjustment made in the [\ion{C}{2}] flux to thus remove
the ionized gas contribution is roughly 50\%. This estimate agrees
with a similar estimate by Petuchowski \& Bennett (1993) for the
contribution to the Milky Way [\ion{C}{2}] by the warm ionized medium. This
is not surprising since we have used Galactic abundance ratios, and
the average $\cn$ in this extragalactic sample agrees with the Milky
Way value. We believe that while statistically this may be a
reasonable treatment to estimate the ionized gas contribution to
[\ion{C}{2}], it may be inaccurate for at least some individual galaxies
because of variations in C/N abundance ratios and in the fraction of
diffuse and dense ionized gas. In Section 6.1, we check for and
discuss the accuracy of estimating [\ion{C}{2}]$_c$.

\subsection{[\ion{O}{1}]/[\ion{C}{2}] compared to PDR models}

\begin{figure}[!ht]
\epsscale{1.0}
\plottwo{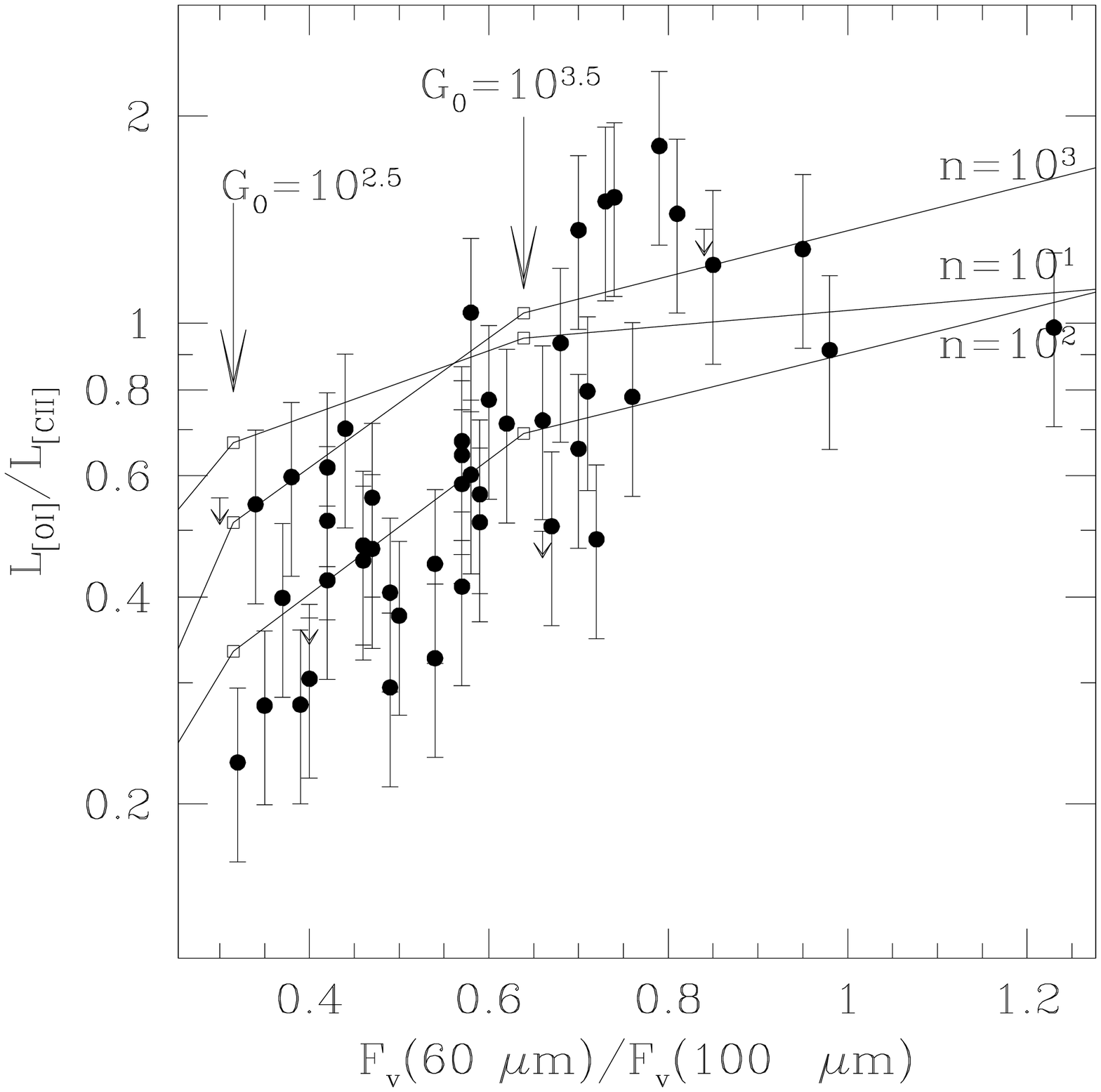}{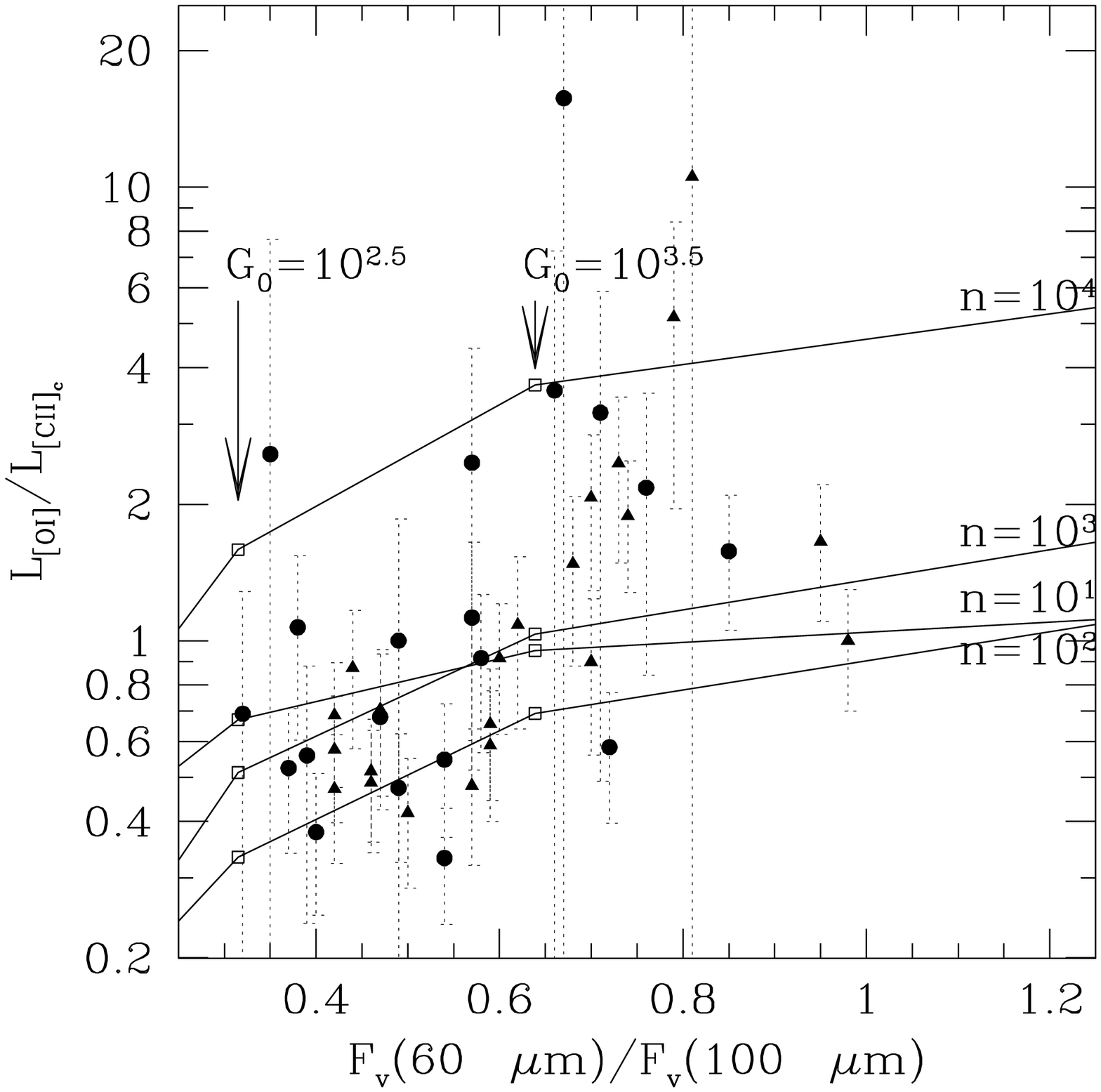}
\caption{(a) The observed correlation between the ratio of the two PDR
cooling lines $\ocrat$ and the dust temperature as indicated
by $\r61$ is shown by the solid dots and upper limits. This is
compared to theoretical PDR models (Kaufman et al. 1999). While
the models run parallel to the trends seen, $\ocrat$ in
some models is too high for about half the sample. (b) The same
comparison is made, except we attempt to subtract the contribution to
[\ion{C}{2}] luminosity from ionized gas by scaling with [NII] luminosity. The solid
points are where [NII] was detected, and the triangles are where we
used the 2-$\sigma$ upper limits on the [NII] measurements. We see
that the scatter in the correlation increases because of the
uncertainties in [NII] observations and the scaling factor used to
subtract the [\ion{C}{2}] emission due to ionized gas (cf section 5.1).  The
correlation between $\ocrat$ and $\r61$ persists, and the PDR models better fit the data. }
\end{figure}

In section 4.3, we noted a strong correlation between the ratio of the
two PDR cooling lines [\ion{O}{1}](63 $\mum$)/[\ion{C}{2}](158 $\mum$) and
far-infrared colors, $\r61$. The increase in $\ocrat$ is
due to the drop in $\llcii$ compared with $\lfir$ 
and not due to an increase in $\orat$. 

Qualitatively, this trend can be explained by postulating that both
the increase in $\ocrat$ and $\r61$ come about in PDRs as a result of
increase in the average FUV flux $G_0$. The overall decrease of
$\rcof$ (Figure 7) means that $G_0/n$ is also increasing. In PDR
models, increasing $G_0$ leads to an increase in the grain temperature
and therefore an increase in $\r61$. Increasing $G_0$ also raises the
gas temperature which raises the ratio $\ocrat$.  Figure 9 shows a
quantitative comparison between the observations and the PDR models of
Kaufman et al. (1999).  While the models run parallel to the trends
seen, the observed $\ocrat$ is too high compared to the models. Some
of this could be due to [\ion{C}{2}] flux from ionized gas.

When the contribution from ionized gas is subtracted using [NII] (122
$\mum$) line measurements as discussed in the last section, the models
and the data show better agreement. In addition, by using the
corrected, [\ion{C}{2}]$_c$, in $\ocrat$, we see that the increase
in $\ocrat$ is due to an increase in both FUV flux.
Comparison with PDR models (Figure 9b) shows that for $n=10-10^3 \cc$
the increase in $\ocrat$ is due to an increase in $G_0$, but for
$n>10^3 \cc$ the $\ocrat$ increases due to an increase in both $G_0$
and $n$.  The critical density for [\ion{C}{2}] is $3 \times 10^3\,\cc$,
while for [\ion{O}{1}] it is $5 \times 10^5\ (T/300)\,\cc$. It is interesting,
then, that $\ocrat$ shows a good correlation with the FIR color, which
does not depend on $n$ but on $G_0$ alone. In the next section we will
present more evidence that $G_0$ and $n$ tend to increase together.

\section{Physical Conditions in the PDRs}

After subtracting an estimated fraction of [\ion{C}{2}] line flux that arises
in ionized gas, we are left with [\ion{C}{2}] line flux arising from PDRs. To
derive the  physical conditions of PDRs we compare the observed
line ratios and line to continuum ratios with 
a grid of recent detailed PDR models by Kaufman et al. (1999) which take into 
account the chemistry, heating and cooling in PDRs.  
These models calculate the line
emission in [\ion{C}{2}](158$\mum$), [\ion{O}{1}] (63 $\mum$ and 145 $\mum$), and dust
continuum emission for a plane slab of gas illuminated from one side
by FUV.  Gas heating is dominated by photoelectrons ejected from
classical (i.e. big, spherical) grains, and from PAHs following the
treatment by Bakes and Tielens (1994). Including the effects of PAH
photoelectric emission raises the PDR surface temperatures by as much as
a factor of 3 compared to the previous PDR models (e.g. WTH).

The comparison of measured fluxes of [\ion{C}{2}] (158$\mum$) and [\ion{O}{1}] (63 $\mum$) lines to
the PDR model predictions allows us to infer the
physical conditions in the PDRs, primarily gas density $n$ and FUV flux $G_0$. 
This method may not give the right answers if: (1) the
various lines and the continuum arise from different
locations in the galaxies; (2) if there is self-absorption of the
lines from cold gas outside the PDRs where they originate.  [\ion{O}{1}] (63
$\mum$) is most susceptible to this, and we try to correct for it in
the models as described below. 

The PDR models are used to calculate the emergent flux from the front
face of a plane parallel slab of gas illuminated from one side. A
galaxy has  many PDRs at all orientations, and 
optical depth effects are non-negligible.  In the approximation that 
most dense PDRs are the shells of molecular clouds, and that the [\ion{C}{2}]
line and dust continuum emission are optically thin whereas [\ion{O}{1}](63
$\mum$) is optically thick, [\ion{O}{1}] is
seen only from the front side of each cloud and [\ion{C}{2}] and FIR from
both the front and the back sides.  The velocity dispersion from cloud 
to cloud however allows most [\ion{O}{1}] (63$\mum$) photons that have escaped 
their parent cloud to escape the galaxy entirely.
This scenario implies that we should observe only half of the [\ion{O}{1}] flux and all
of the [\ion{C}{2}] and dust continuum flux expected from PDR models. 
This adjustment has been applied hereafter to the PDR model predictions.

Figure 10(a) shows the measured ratios $\rm L_{\rm [OI]}/\rm L_{\rm
[CII]_c}$ and ($\rm L_{\rm [OI]}+\rm L_{\rm [CII]_c})/\rm L_{\rm TIR}$
compared with the values obtained from a grid of PDR models with various values of G$_0$
and $n$. With just these two measured ratios there are two possible
regimes of G$_0$ and $n$ which reproduce the observed values.  One of
the regimes is shown in Figure 10(a), while the other is a high $n$ ($10^4\,\cc
<n< 10^5\,\cc$), low G$_0$ ($G_0 \simeq 1$) regime. To distinguish
between the two, we need to reliably measure the ratio $\rm L_{\rm
[OI]145}/\rm L_{\rm [OI]63}$. This measurement is not available for
most of the galaxies in this sample, [\ion{O}{1}](145 $\mum$) being too
faint. However the high $n$, low G$_0$ solution does not reproduce the
line fluxes measured, i.e. the fluxes in the [\ion{C}{2}] line fall short by
about 2 orders of magnitude when $10^4 <n< 10^5$ and $G_0 \simeq 1$
compared to the observed values. This is not to say that such high
density, low radiation environment (e.g. molecular clouds) do not
exist in galaxies, they simply do not contribute much to the observed
fluxes seen in [\ion{O}{1}], [\ion{C}{2}] lines and the FIR continuum.

\begin{figure}[!htb]
\epsscale{1.0}
\plottwo{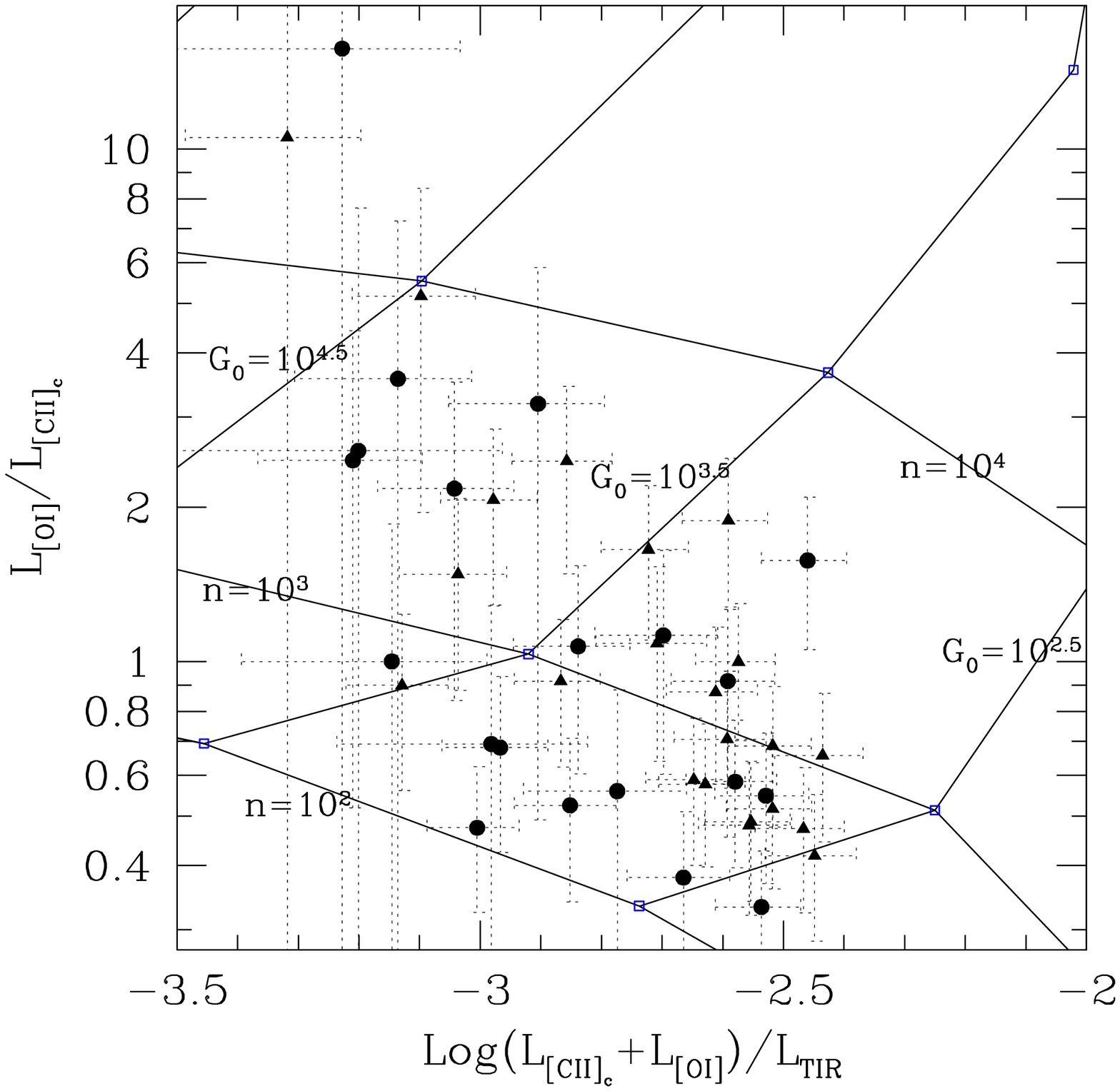}{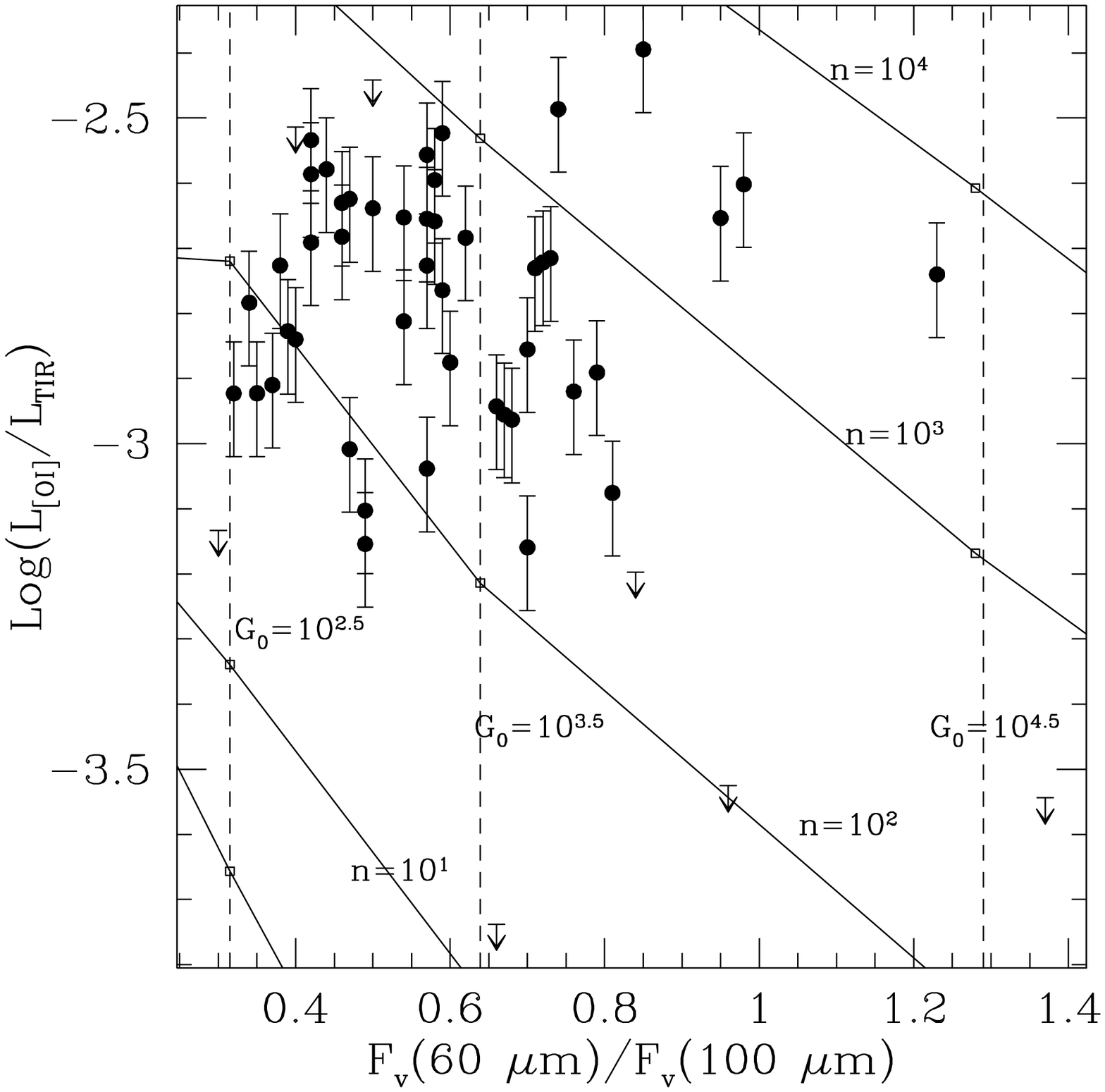}
\caption{(a) The measured ratios $\rm L_{\rm [OI]}/\rm L_{\rm
[CII]_c}$ and $(\rm L_{\rm [OI]}+\rm L_{\rm [CII]_c})/\rm L_{\rm TIR}$
(circles and triangles) are overplotted with a grid of PDR model
values of the same. Solid lines are contours of constant gas density
$n$ and the dotted lines are contours of constant FUV radiation flux $G_0$. Two
regimes of $G_0$ and $n$ reproduce the observed ratios. Only one is
shown here for clarity. The other regime is ruled out as it would not
produce the observed [\ion{C}{2}] and [\ion{O}{1}] fluxes. The observed points are as
in Figure 10: filled circles represent values where [\ion{C}{2}] was
corrected for the contribution from ionized gas using measured values of [NII], and
triangles are where upper limits for [NII] were used. (b) The measured
ratios of $L_{[OI]}/L_{TIR}$ and $\r61$ are plotted along with a grid
of $L_{[OI]}/L_{TIR}$ and $\r61$ values calculated according to PDR
models of Kaufman et al. (1999) and dust models of Hollenbach, Takahashi
\& Tielens (1991).}
\end{figure}

\subsection{G$_0$ and n from [\ion{O}{1}]/TIR values}

We can also use the $L_{[OI]}/L_{TIR}$ ratio to derive G$_0$ and
n by comparing the observed $L_{[OI]}/L_{TIR}$ vs. $\r61$ with models. Such
a comparison, shown in Figure~11(b), using models by Kaufman et al. (1999), 
suggests that the
constancy of [\ion{O}{1}]/FIR (or TIR) against variations in $\r61$ implies a
simultaneous increase in both gas density $n$ and FUV flux $G_0$. The model 
$\r61$ ratio for a given UV flux
G$_0$ is calculated following the treatment by Hollenbach, Takahashi
\& Tielens (1991). The 60 and 100 $\mum$ fluxes are calculated assuming
classical grains in thermal equilibrium. For low G$_0$, the emission
from classical grains is from the Wein side of the blackbody curve, and the
$\r61$ is set by stochastically heated grains. This has the effect of
yielding essentially the same $\r61$ for G$_0=1-10^{2.5}$.  The other
uncertainty in using $L_{[OI]}/L_{TIR}$ to derive G$_0$ and n is
that [\ion{O}{1}](63 $\mum$) may be optically thick; we have accounted for this by
reducing $L_{[OI]}/L_{TIR}$ from the models by a factor of two, as before.
An advantage to this disagnostic however is that we 
avoid uncertainties associated with correcting for the [\ion{C}{2}] flux
from ionized regions.

Figure 11 compares the $G_0$ and n values obtained using each set of
observables, $\rm L_{\rm [OI]}/\rm L_{\rm [CII]_c}$ vs ($\rm L_{\rm
[OI]}+\rm L_{\rm [CII]_c})/\rm L_{\rm TIR}$ and $L_{[OI]}/L_{TIR}$ vs
$\r61$. The difference in log(G$_0$) obtained by these two methods
$\Delta Log(G_0)$ shows a mean of 0.08 and $\Delta log(n)$ shows a mean of
0.06. Thus there is no indication that one method yields
systematically higher or lower values of G$_0$ or $n$. From the
absolute values of $\Delta G_0$ and $\Delta n$, we infer that for half
the galaxies $G_0$ and $n$ values from the two methods agree to better
than a factor of 2 and 1.5, respectively. This difference may be taken
as an estimate of uncertainty in the derived values, which is
sensitive to some measurement errors (one method uses only the [\ion{O}{1}]
line while the other uses [\ion{C}{2}], [\ion{O}{1}] and [NII]) and to some sources
of systematic errors (for example, subtraction of the ionized gas
contribution to [\ion{C}{2}] flux), but not to all possible systematics.  For
instance, this estimate would not be sensitive to uncertainties in the
abundances used in the models, or to the assumption that about half
the dust heating is due to FUV photons, or to the fact that we measure
luminosity weighted averages of line and continuum contributions
emanating from many parcels of ISM with different physical conditions.

\begin{figure}[!htb]
\epsscale{0.5}
\plotone{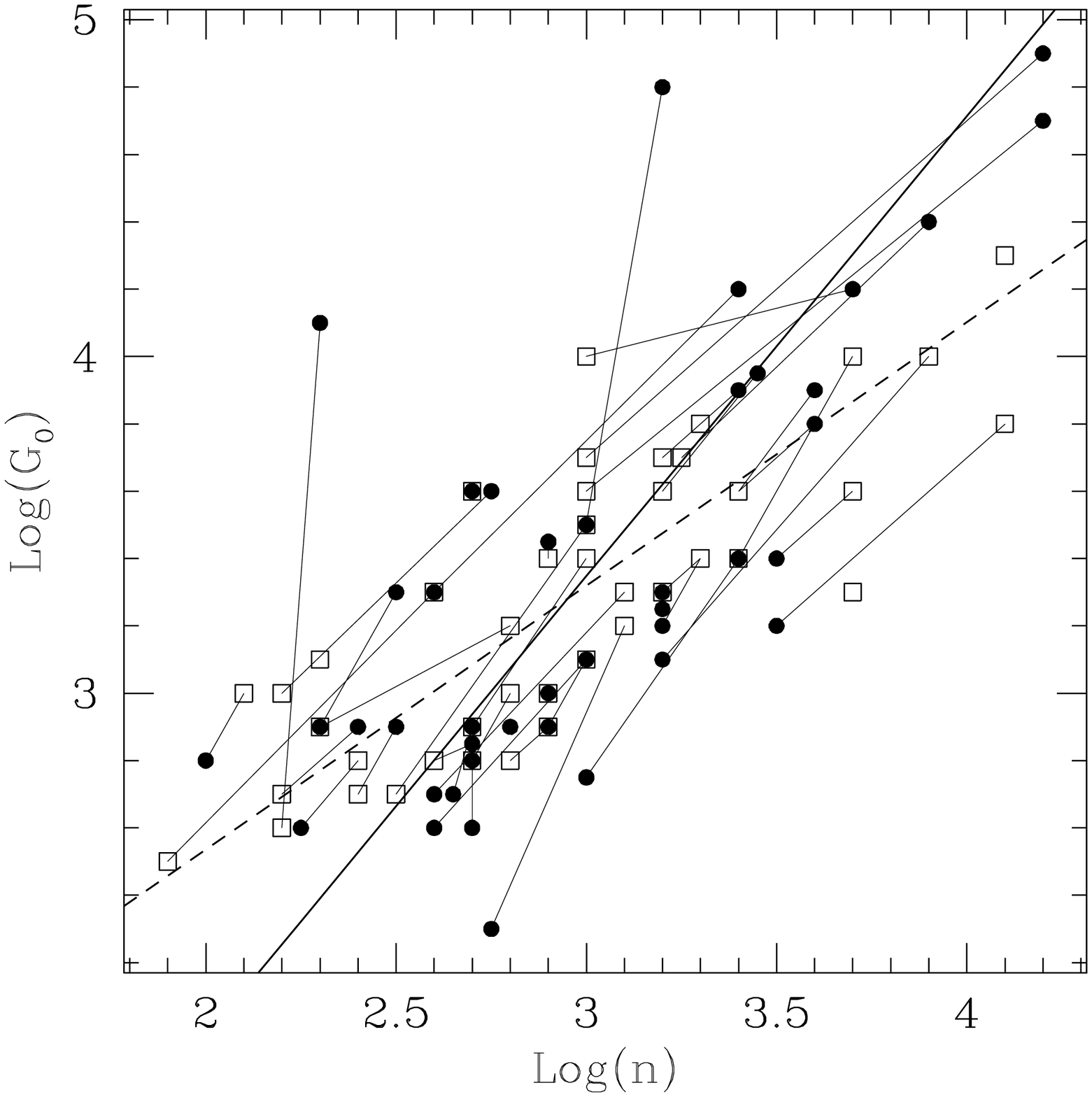}
\caption{This figure shows the G$_0$ and n solution for  galaxies
based on comparison of data and PDR models shown in figure 11. The
filled circles are G$_0$ and n values estimated from $\rm L_{\rm
[OI]}/\rm L_{\rm [CII]_c}$ and $(\rm L_{\rm [OI]}+\rm L_{\rm
[CII]_c})/\rm L_{\rm TIR}$; and the open squares are the G$_0$ and n
values derived from $L_{[OI]}/L_{TIR}$ and $\r61$. The G$_0$ and $n$
values derived from the two methods are connected for each source to
give an estimate of the uncertainty in the parameters. A least squares
fit is made to both sets of $G_0$ and $n$ values assuming equal
error in both axes.  The best fit slopes are 1.4 and 1.3 respectively,
i.e. G$_0 \propto n^{\alpha}$, with $\alpha=1.3-1.4$, which is
consistent with the emission coming from PDRs surrounding ionization-bounded 
expanding \ion{H}{2} regions.}
\end{figure}

\subsection{Discussion and interpretation}

The average physical conditions in this sample of normal galaxies are
FUV flux $G_0=10^{2}-10^{4.5}$ and gas densities
$n=10^2-10^{4.5}\,\cc$. The derived values of $G_0$ and $n$ follow a
trend with G$_0 \propto n^{\alpha}$, $\alpha=1.3$ or 1.4
(Figure 12). This may be evidence for a Schmidt law on fairly local
scales in the galaxies, where high gas densities correlate with high
star formation rates (Schmidt 1959) and agrees with the recent
determinations by Kennicutt (1998) where the surface densities of
star-formation and gas scale by an exponent N=1.4: $\Sigma_{SFR}
\propto \Sigma_{gas}^{1.4}$. The observed scaling between G$_0$ and n
may be also dictated simply by geometry if the FIR line and continuum
emission is dominated by regions near young stars. From simple
Str\"{o}mgren sphere calculations (cf. Spitzer 1978) we can derive
that the FUV flux at the neutral surface just outside the
Str\"{o}mgren sphere should scale as G$_0 \propto n^{4/3}$, which is
consistent (within errors) to the scaling seen in Figure 11.

We favor the latter interpretation primarily because it is the more
direct one. The Schmidt law applies to surface density or a density
which is averaged over the entire galactic disk thickness. It reflects
the physics that on a {\it global} scale, the star formation rate may
scale as the mass (density) or, if cloud collisions are important,
for example, on a higher power of the mass (density). However, the
high values of $G_0$ and $n$ which we derive make it clear that we are
probing a very local phenomenon. We are not  probing the rate
of star formation in as much as the typical distance from an OB star or OB
association to a PDR which absorbs its FUV photons. O and early B
stars form in dense cores within giant molecular clouds (GMCs), whose
diameters are of order 10--50 pc. For a fraction of their lives, the
OB stars are embedded in the clouds, surrounded by very dense and
compact \ion{H}{2} regions and PDRs illuminated by intense UV fluxes. Later,
the expanding \ion{H}{2} regions break through the surfaces of the clouds,
champagne flows are created, and eventually the OB stars lie outside,
but close, to their natal GMCs. To produce a flux of $G_0\approx 1000$
on the surface of that cloud, an O7 star must be only about 1 pc
distant. With a relative speed of 1 $\kms$ with respect to the cloud,
an O star travels only 1 pc in $10^6$ years, a substantial portion of
its lifetime. Therefore, the FUV photons from OB stars tend to be
absorbed in two very different environments. A substantial fraction
(McKee \& Williams 1997 estimate $\ge 30\%$ for the Lyman continuum
photons) are absorbed in their immediate vicinity - in the GMC of
their birth. The rest escape from the natal cloud and travel through
the diffuse ISM to be absorbed primarily by diffuse gas at distances
of order 100--300 pc, with much lower values of $G_0$ and $n$ ($\sim$2
and 100 $\cc$ in the local ISM of the Milky Way). These two components
are consistent with the decomposition of the FIR dust emission into
``active'' and ``cirrus'' components by Helou (1986).  [\ion{C}{2}]
(158$\mum$) can have significant contributions from the DIG, the
diffuse neutral gas, and the locally irradiated PDRs of the natal
GMCs. [\ion{O}{1}] (63$\mum$) is dominated by emission from the PDRs of the
natal GMCs. The PDR modeling procedure provides an average of these
two components, weighted to the high $G_0$ and high $n$ GMC component.
The simple Str\"{o}mgren sphere scaling G$_0 \propto n^{4/3}$ applies
directly to this GMC component, and should therefore manifest itself
in the integrated fluxes as seen in Figure 12.

The derived PDR pressures \footnote{The PDR pressure is calculated as
$P=nkT$ where gas density $n$ and temperature $T$ are derived from PDR
models (see Kaufman et al. 1999 for details)} also lend credence to the
view that the $G_0\propto n^{1.4}$ correlation arises from the
correlation of $n$ and $G_0$ in an expanding \ion{H}{2} region surrounded by
a PDR. In such a picture, the \ion{H}{2} region thermal pressure should
approximately equal the PDR thermal pressure. The thermal pressure in
the \ion{H}{2} regions can be estimated from the electron density derived
from [\ion{O}{3}]
(88$\mum$)/[\ion{O}{3}]
(52$\mum$) ratio if both lines are
observed (Rubin et al. 1994). This comparison was made for the three
galaxies in our sample with measurements of the [\ion{O}{3}]
 lines. The
pressures derived for \ion{H}{2} regions are tabulated in Table 7. The \ion{H}{2}
region and PDR pressures agree to roughly a factor of two in all cases
(assuming $T_e=8000 K$).

\begin{figure}[!htb]
 \epsscale{0.5}
\plotone{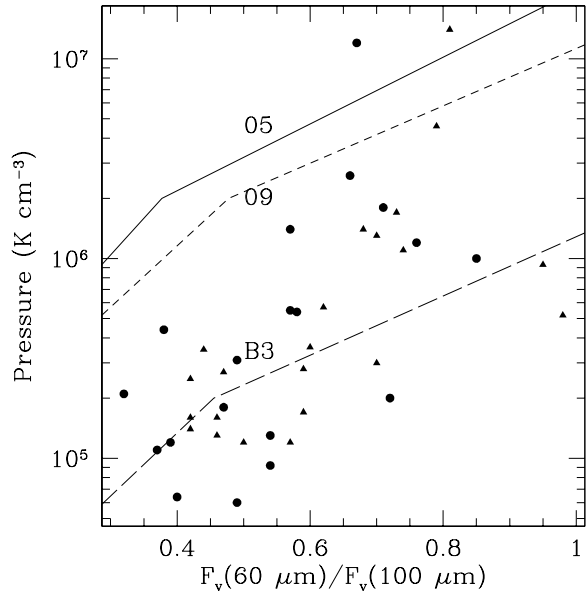}
\caption{ The average pressure near star forming regions in the galaxies is plotted against 
the FIR colors. The pressure is a product of the gas densities 
n and temperatures T derived from the PDR models. Both $n$ and T increase 
with $\r61$, but the increase in $n$ is more dramatic and is responsible
for the correlation between pressure and $\r61$. These pressures likely do not reflect the average thermal pressure of the ISM, but do represent the pressures in PDRs surrounding \ion{H}{2} regions associated with O and B stars. Curves are shown for the pressure 
in PDRs surrounding \ion{H}{2} regions formed by stars of different spectral types (see text).}
\end{figure}

The derived temperatures at the PDR surfaces range from 270-900 K, and
the pressures range from $ 6 \times 10^4 - 1.5 \times 10^7\,{\rm K}\,
\cc$ (Table 6). The lower value of the pressure range is roughly twice
the local solar neighborhood value and the upper end is comparable to
pressures in \ion{H}{2} regions in starburst galaxies (Heckman, Armus \&
Miley 1990) which also corresponds to the pressure and surface
brightness at which starbursts saturate (Meurer et al. 1997).

Figure 12 shows that the thermal gas pressures increase with grain
temperatures, or the flux of photons incident upon the absorbing
grains.  This is mainly due to the $\lo/\llcii$ vs $\r61$ correlation
in Figure 5. Higher values of $\lo/\llcii$ imply higher G$_0$ and gas densities,
and therefore higher pressures. The temperature of the gas also
increases with $\r61$ and contributes to the increase in pressure,
but the increase in the gas density n is the more dramatic of the
two. This too is explicable in terms of the local absorption of
photons by the expanding \ion{H}{2} regions and associated PDRs in the natal
GMCs of OB stars. Whether grains absorb most of the flux in the \ion{H}{2}
region or in the neighboring PDR, the photons will be absorbed at
roughly $R_S$, the Str\"omgren radius for that star. If $\phi_i$ is
the Lyman continuum photon luminosity of the star, then the photon
flux is proportional to $\phi_i/R_S^2\propto n_e^{4/3}\propto P_{\rm
HII}^{4/3} = P_{\rm PDR}^{4/3}$, assuming pressure equilibrium between
the \ion{H}{2} gas and the PDR regions associated with them. Since the grain
temperature scales roughly as the photon flux to the $1/5^{\rm th}$
power (Hollenbach, Takahashi \& Tielens 1991), $T_{gr}\propto P_{\rm
PDR}^{4/15}$.

We show a simple model to help explain the observed correlation
between the FUV field strength, $G_0$, and the PDR pressure, P. If we
assume that PDR gas is illuminated by hot stars in star clusters, then
there is an approximate natural scaling between the FUV flux from
stars in the cluster, the \ion{H}{2} region density, and the pressure at the
\ion{H}{2} region/PDR interface. Assuming that all of the FUV flux emerges
from one star of a given spectral type which has a bolometric
luminosity, $L_{bol}$ and which produces $S=10^{49}S_{49}$ photons
s$^{-1}$ of ionizing photons, the size of the surrounding \ion{H}{2} region,
$R_s$ is set by the number of ionizing photons and the electron
density in the \ion{H}{2} region. Assuming further an electron temperature,
$T_e=10^4\,$K, the combination $n_eT_e$ sets the \ion{H}{2} region pressure.
Then, for a given bolometric stellar flux, the value of $G_0$ at the
PDR surface is determined by $G_0=L_{bol}/4\pi R_s^2$. If there is a
pressure equilibrium at the \ion{H}{2} region/PDR interface, then we find a
correlation between $G_0$ and $P$. This is plotted as a correlation
between $\r61$ and $P$ (Figure 13). The calculations use three
different spectral types (O5, O9, and B3) and electron densities
ranging from $10^1$ to $10^8\,\rm cm^{-3}$.  Zero-age main sequence
values of $L_{bol}$ and $S_{49}$ for the various spectral types are
from Parravano (private communication). 

Because the ISO-LWS beam views an entire galaxy for our sample
galaxies, it incorporates many tens to hundreds of thousands of O and
early B stars, with their associated \ion{H}{2} regions and PDRs. As
discussed above, the $G_0$, $n$ and $P$  we derive for a given
galaxy represent an average value, weighted to the dense natal GMCs
which lie close to the OB stars. The range of the $G_0$, $n$ and $P$ 
 for the different galaxies can reflect several interesting
differences in their star formation processes and histories. If global
star formation in normal galaxies occurs in bursts, rather than
continuously, then those galaxies which are observed shortly after
their bursts ($\lesssim$ few million years) will show high $G_0$, $n$
and $P$ because the OB stars will not have had time to travel very far
 from their natal clouds.  This hypothesis, however, requires the
star formation bursts to be of significant amplitude, resulting in
sustantially different instantaneous mass distributions of ionizing
stars among galaxies.

The differences in $G_0$, $n$ and $P$ may instead reflect differences
from galaxy to galaxy in the GMCs which form the OB stars. For
example, larger GMCs may keep their OB stars embedded for a longer
fraction of their lifetime, resulting in higher average $G_0$ and
$n$. We note that $n$ is the density in the PDR, which can be
considerably higher than the average density in the GMC. On the other
hand, the GMCs could be the same size but denser. In this scenario,
the higher density ambient gas would lengthen the embedded phase of
the OB star, and result in higher average $G_0$ and $n$. In either
case, the galaxies with higher derived $G_0$ and $n$ would contain
more massive GMCs, on average.

\section{Summary and Conclusions}

In this paper we have attempted to understand the energetics and the
physical conditions in a statistically representative set of star-forming
normal galaxies by studying the atomic and ionic fine structure lines
in the far-infrared. Such sensitive observations of a large sample were
made possible by having a cryogenically-cooled observatory, ISO
(Kessler et al. 1996), in space. The sample was selected to span a
range in properties such as morphology, FIR colors $\r61$
(indicating dust temperatures), and $\brat$ (indicating
star-formation activity and optical depth). For a randomly drawn sample
of galaxies, many of these parameters are correlated. Care was taken
to span the full range of parameter space as much as possible in this
sample.  Still, there remains some correlations between these
parameters. Galaxies with high $\brat$ also tend to have
warmer FIR colors and tend to be luminous.

Since this sample is more extensive than previous studies, many
effects were seen for the first time. Among the more remarkable was
the non-detection of [\ion{C}{2}] line in two normal galaxies
indicating low $\rat$, down to $2 \times 10^{-4}$ (3-$\sigma$ upper
limits). Lower ratios have been seen in association with Galactic
\ion{H}{2} regions, but such low ratios were unexpected in normal
galaxies where the emission is from a mixture of sources: \ion{H}{2}
regions, WIM, and PDRs. Previous observations of galaxies had found
$\rat$ to vary little, between $10^{-2}$ and $10^{-3}$. This result
may have been biased because of the low sensitivity of the previous
surveys and the selection criteria for the galaxies. DIRBE
observations of the Milky Way also showed a spatially constant ratio
$\rat$=$3 \times 10^{-3}$, except in the Galactic center.  In the
current sample of normal galaxies we see a smooth decline in $\rat$
with increasing dust temperature and star-forming activity in
galaxies. In a sample of 60 normal galaxies, this trend spans a factor
of more than 50 in $\rat$ with [\ion{C}{2}] deficient galaxies at the
hottest and most active end. The anticorrelation between $\rat$ and
$\6f/\f100$ is the strongest, followed by $\rat$ vs. $\lfir/\lb$. The
anticorrelation between $\rat$ and the IR-luminosity is the weakest
and may be a secondary correlation. This is good news for searches for
distant startbursts. The most luminous galaxies are not necessarily
[\ion{C}{2}] deficient: warm galaxies are more likely to show a lower
$\rat$, whereas cool luminous galaxies can be detected.

There have been many explanations for the variations in $\rat$
, but on examining the evidence we favor the scenario where
$\rat$ decreases as the heating of the gas becomes less efficient
in the high $G_0/n$ regime due to charging of dust grains. As the grains
become positively charged the efficiency of photoelectric ejection
decreases. The various lines of evidence supporting this hypothesis are:
\\
(1) $\rat$ deficiency is seen in the more actively star-forming galaxies
with warmer FIR colors.\\
(2) Examination of heating and cooling balance in Arp~220 does not show
any other lines that could be cooling the neutral ISM instead of [\ion{C}{2}]. In our
own sample, the inclusion of [\ion{O}{1}] does not change the low ratio of gas to 
grain heating as indicated by $\rcof$. So far we have failed
to identify other channels, which leads us to believe that heating efficiency
is low in galaxies where where $\rat$ is low.\\
(3) PDR models which include grain charging and the photoelectric effect 
successfully reproduce the trends of decreasing $\rcof$ and $\rat$ 
with $\r61$ (Figure 8).

The [\ion{C}{2}] line flux shows better correlation (and less scatter) with
mid-IR flux from galaxies which is dominated by aromatic features and
emission from transiently heated small grains (Helou et
al. 2001). This indicates that the heating of gas is dominated by such
grains as expected by some theoretical considerations (e.g. Bakes \&
Tielens 1994).

A somewhat less dramatic decrease in $\rat$ is seen in early type
galaxies where this decrease is due to softer radiation
fields. Pierini et al. (1999) also see a decrease in $\rat$ with
decreasing star formation in quiescent galaxies.  In early-type
galaxies with little or no star-formation, $\llcii$ is lower because
UV photons are needed to heat the gas by photoelectrons, whereas both
optical and UV photons can heat the dust. This scenario is
corroborated by high ratios of $\rm L_{\rm CO}/\rm L_{\rm [CII]}$ in some
early-type galaxies (Malhotra et al 2000). At the other morphological
extreme, irregular galaxies show high $\o3f$ (Figure 6) and much of
the [\ion{C}{2}] seen in these galaxies may be originating in the ionized
regions (Hunter et al. 2001). Apart from these effects, we do not see
much dependence on galaxy morphology.

Because carbon has an ionization potential lower than hydrogen, [\ion{C}{2}]
emission can arise in both ionized and predominantly neutral
media. The observed ratio $\ocrat$ is lower than the ratio in PDR
models, suggesting a possible contribution of [\ion{C}{2}] emission from
ionized gas.  In addition, the ratio of [\ion{C}{2}] and [NII] fluxes trace
each other reasonably well in the spatially resolved data in the Milky
Way (Bennett et al. 1994). These two facts suggest that a fair
fraction of [\ion{C}{2}] arises in ionized regions.  It is also significant
that $\ocrat$ shows a remarkably tight correlation with $\r61$ which
is as expected from PDR models, suggesting that a significant fraction
of [\ion{C}{2}] arises from PDRs.  Thus the PDRs and diffuse ionized gas are
associated with each other.
 
We estimate a scaling between [NII](122 $\mum$) and [\ion{C}{2}] (158
$\mum$) from diffuse and dense ionized gas and use that scaling to
estimate the [\ion{C}{2}] luminosity from neutral regions $\rm L_{\rm
[CII]_c}$. The main uncertainty in this calculation comes from the
relative abundances of C and N. Additional uncertainty is contributed by the
unconstrained fraction of 
the [\ion{C}{2}] and [NII] emission arising from dense vs. diffuse ionized
medium, since the scaling of $\cn$ is different for the two.

By comparing the observed ratios of the [\ion{O}{1}](63$\mum$) and the
[CII]$_c$ lines from PDRs and the ratio of line to the total FIR
continuum from dust with a grid of PDR models (Kaufman et al. 1999),
we derive the average $G_0$ to be in the range $10^2 - 10^{4.5}$, and gas
densities $n$ in the range $10^2 - 10^{4.5}\,\cc$ for this sample of
galaxies. The FUV flux $G_0$ and gas densities $n$ correlate with each
other and $G_0$ increases roughly proportional to $n^{1.4}$ over
about two orders of magnitude. We also derive $G_0$ and $n$ from the
observed values of $\orat$ vs $\r61$. Averaged over the sample there
is no systematic difference between $G_0$ and $n$ derived by the two
methods. Half the galaxies show better than a factor of two agreement
in $G_0$ and $n$ derived from the two methods. 

The correlation between $G_0$ and $n$ is explained by assuming that a
significant portion of the PDR emission comes from PDRs surrounding
expanding \ion{H}{2} regions. The range in values of $G_0$ and $n$ derived
for this sample suggests that the GMCs in which OB stars are born have
different average properties from galaxy to galaxy. We suggest that
the galaxies with higher $G_0$ and $n$ have more massive GMCs.

\acknowledgements 

We thank the anonymous referee for valuable comments that improved
this paper. This work was supported by ISO data analysis funding from
NASA, and carried out at IPAC and the JPL of the California Institute
of Technology.  SM's research funding is provided by NASA through
Hubble Fellowship grant \# HF-01111.01-98A from the Space Telescope
Science Institute, which is operated by the Association of
Universities for Research in Astronomy, Inc., under NASA contract
NAS5-26555. ISO is an ESA project with instruments funded by ESA
Member States (especially the PI countries: France, Germany, the
Netherlands and the United Kingdom), and with the participation of
ISAS and NASA. This research has made use of the NASA/IPAC
Extragalactic Database (NED) which is operated by the JPL, California
Institute of Technology, under contract with the NASA.

\appendix

\section {Sample selection}

This sample was constructed to study star-formation and the ISM in
normal galaxies. Here normal galaxies are deined to have the energy
production in the galaxy come from star-formation and not active
nuclei.

The distant galaxies were selected to span a range of galaxy
properties (Dale et al. 2001):\\
(1) Morphology: The morphology of the galaxies in this sample ranges
from Irr through E (Table 5). The galaxies are uniformly distributed 
across the range of morphological types (see Dale et al. 2000). \\
(2) Far-infrared luminosity: The FIR luminosity of galaxies in this
sample span the range Log($\lfir/\lsun$)=7.7-11.2. Since this is a
study of normal galaxies, we avoided ultra-luminous galaxies which
might harbor hidden AGNs.\\
(3) $\r61$: The ratio of FIR fluxes in the IRAS filters at 60 and 100
$\mum$ indicates the average dust temperature in the galaxies. The
galaxies in the sample cover the range $\r61=0.3-1.37$.\\
(4) $\lfir/\lb$: The FIR to B-band flux ratio is a measure of
star-formation activity and of dust optical depth. For low values
of $\brat$ one can approximate $\lfir$ as surrogate of extincted UV
and hence current star-formation, while $\lb$ represents relatively
old stars. High values of $\brat$ indicate high star-formation rates
as well as high optical depth in dust.

Care was taken during sample selection to sample the parameter space
in these four parameters: morphology, IR luminosity, $\6f/\f100$ and
$\lfir/\lb$, as much as possible. For a randomly selected sample, the
quantities IR luminosity, $\6f/\f100$ and $\lfir/\lb$, correlate
with each other. We have tried to reduce these correlations by picking
galaxies that sample the parameter space as uniformly as
possible. Still, correlations between IR luminosity, $\6f/\f100$ and
$\lfir/\lb$, remain and are illustrated in Figure 14. 

\begin{figure}[htb]
\epsscale{1.0}
\plottwo{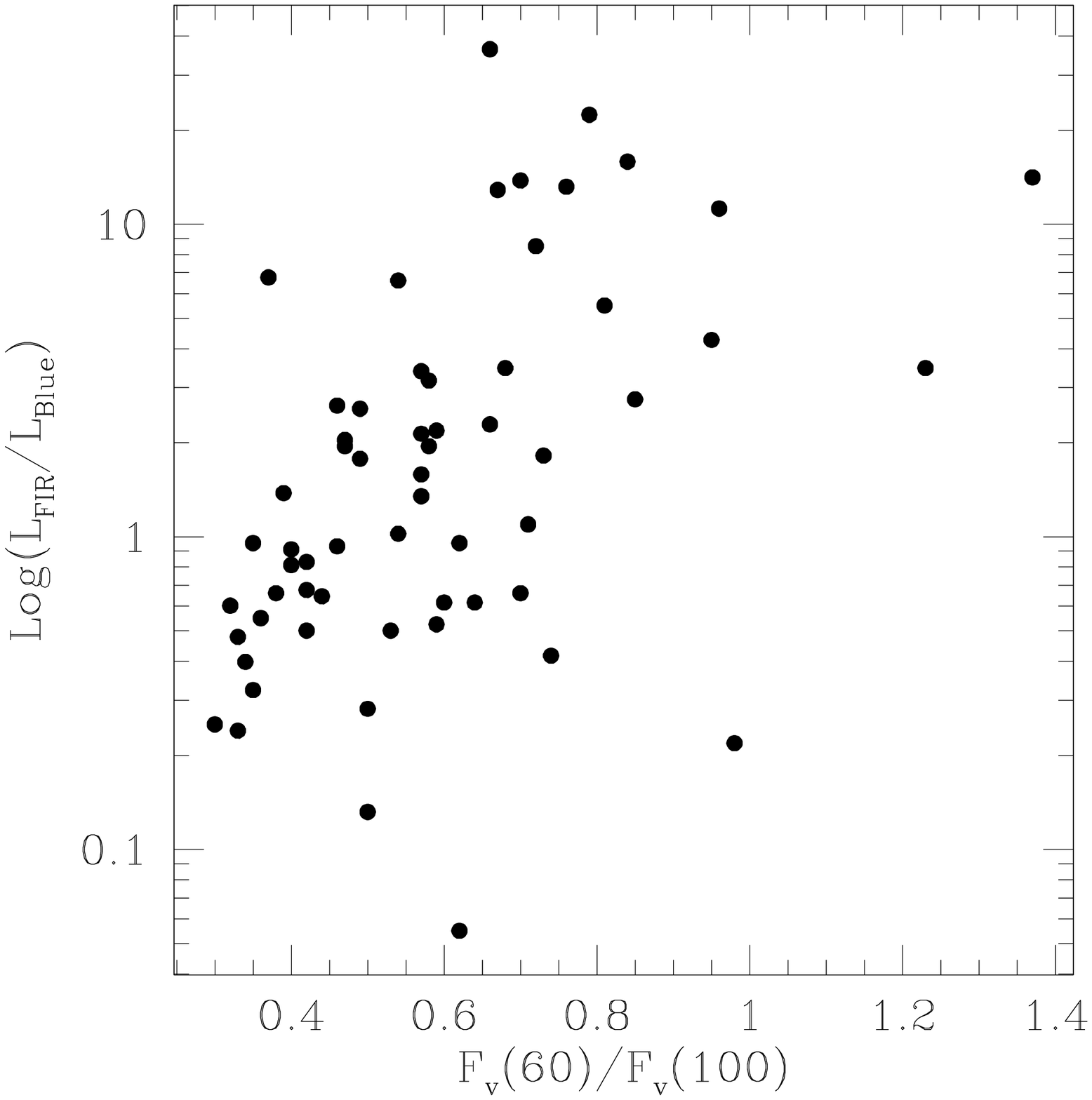}{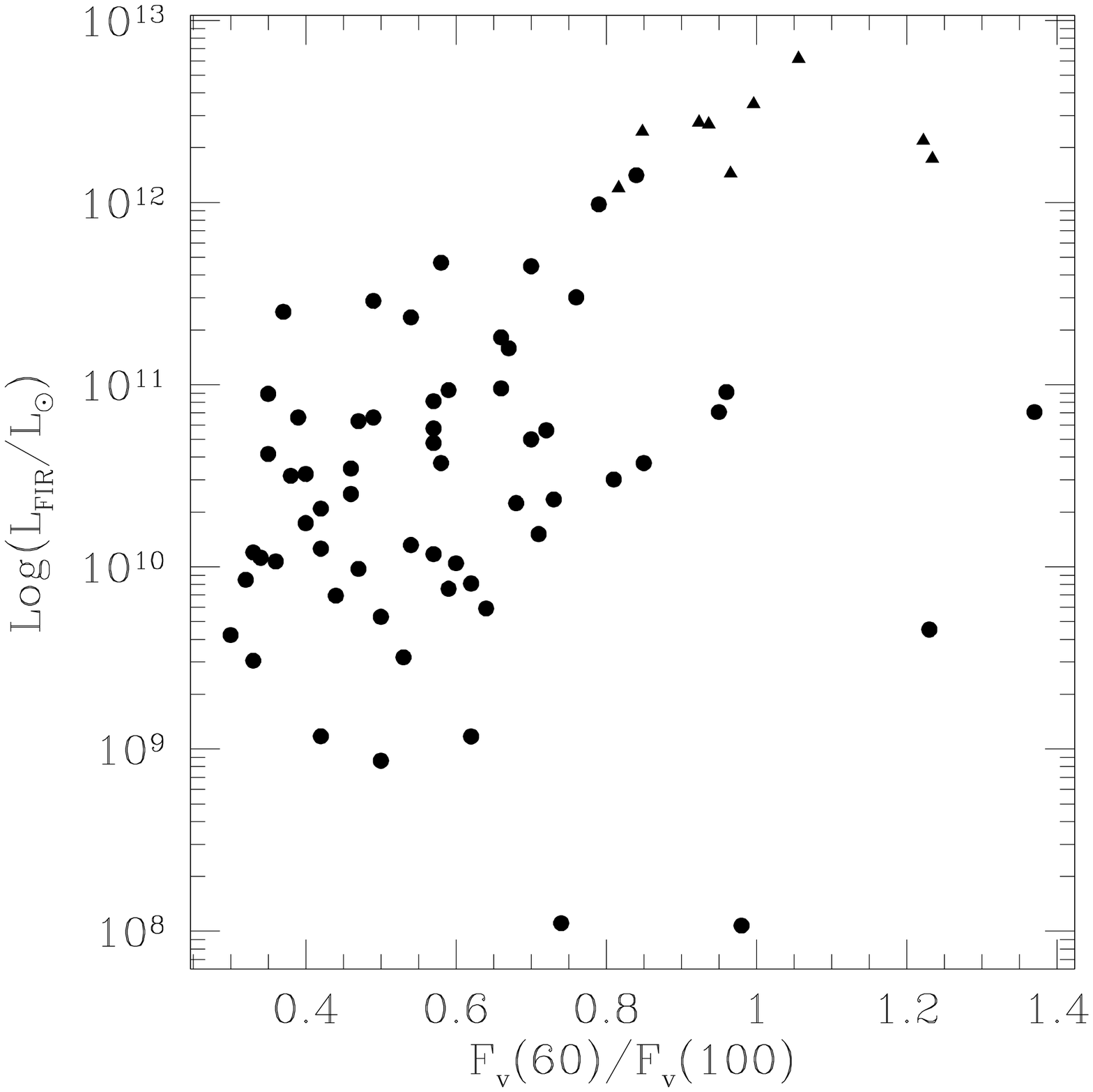}
\caption{These two figures show how the galaxies discussed in this
paper span the parameter space along the parameters $\r61$, $\brat$
and infrared luminosity. In figure 13(a), a correlation between $\r61$
and $\brat$ is seen.  This correlation is 4.5 $\sigma$
significant. Figure 13(b) also shows in triangles the luminous and ultraluminous
galaxies from the sample of Luhman et al. (1998). The correlation between 
$\r61$ and $L_{FIR}$ is 4.1$\sigma$ significant.}
\end{figure}
\begin{table*}[htb]{}
\scriptsize
\caption[ ]{The sample}
\begin{flushleft}
\begin{tabular}{lccccl}
\hline
Galaxy    & Morphology & $F_\nu (60 \mu m Jy)$ & $\frac{F_\nu (60 \mu m)}{F_\nu (100 \mu m)}$ & $\frac{L_{FIR}}{L_{B}}$& FIR (10$^{-14} W/m^2$)\cr
\hline
  NGC 0278  & SAB(rs)b  & 25.05  &  0.54  &  0.01  & 139.9 \cr
  NGC 520   &Irr  &  31.1  &  0.66  &  0.36  & 160.5\cr
  NGC 0693  & I0: sp  &  6.73  &  0.57  &  0.13  & 36.75\cr
  NGC 0695  & IB?(s)m: pec  &  7.87  &  0.58  &0.5  & 42.68\cr
 UGC 01449  & SBm; pec:  &  4.96  &  0.59  &  0.34  & 26.72\cr
 MCG-03-06-01  &  SB0 pec: &  4.41  &  1.23  &  0.54  & 18.85\cr
  NGC0986  & (R\'\_1)SB(rs)b  & 25.14  &  0.49  &  0.25  & 146.4\cr
  NGC1022  &  (R\')SB(s)a  & 19.83  &  0.73  &  0.26  & 98.69\cr
  NGC1052  & E4;Liner;Sy2  &  0.93  &  0.62  & -1.26  & 4.913\cr
 UGC02238  & Pec  &8.4  &  0.54  &  0.82  & 46.91\cr
  NGC1155  & (R\')SAB(s)0o: pec  &  2.89  &  0.58  &  0.29  & 15.67\cr
  NGC1156  &IB(s)m  &  5.24  &0.5  & -0.55  & 30.24\cr
  NGC1222  &  S0-pec:  & 13.07  &  0.85  &  0.44  & 61.86\cr
 UGC02519  & SAB?(s:)cd III:  &  2.98  &0.4  & -0.04  & 19.07\cr
  NGC1266  & (R\')SB(rs)0  pec;Liner  & 13.32  &  0.81  &  0.74  & 64.02\cr
  NGC1317  & (R\')SAB(rl)0/a  &  3.52  &  0.34  &  -0.4  & 24.49\cr
  NGC1326  & (R-1)SB(rl)0/a;Liner  &  8.17  &  0.59  & -0.28  & 44.01\cr
  NGC1385  &  SB(s)cd  &  17.3  &  0.46  & -0.03  & 103.6\cr
 UGC02855  & SB(s)cd II-III  & 42.39  &  0.47  &  0.31  & 251.4\cr
  NGC1482  & SA0+;pec;sp  & 33.45  &  0.72  &  0.93  & 167.3\cr
  NGC1546  &  SA?a pec  &  7.21  &  0.32  & -0.22  & 51.83\cr
  NGC1569  &  IBm;Sy1  & 54.25  &  0.98  & -0.66  & 246.1\cr
  NGC2388  &  SA(s)b: pec  & 17.01  &  0.67  &  1.11  & 87.29\cr
 ESO317-G023  & (R\'-1)SB(rs)a  &  13.5  &  0.57  &  0.53  & 73.73\cr
 IRASF10565+2  & Pec  & 12.08  &  0.79  &  1.35  & 58.54\cr
  NGC3583  &SB(s)b  &  7.08  &  0.38  & -0.18  & 46.49\cr
  NGC3620  & (R\'\_1)SB(s)ab  &  46.8  &0.7  & -0.18  & 236.4\cr
  NGC3683  &  SB(s)c?  & 13.61  &  0.46  &  0.42  & 81.52\cr
  NGC3705  & SAB(r)ab  &  3.72  &  0.33  & -0.62  &  26.3\cr
  NGC3885  & SAB(r:)0/a:  & 11.66  &  0.71  &  0.04  &  58.6\cr
  NGC3949  & SA(s)bc:  & 11.28  &  0.44  & -0.19  & 68.97\cr
  NGC4027  &  SB(s)dm  & 11.61  &  0.42  & -0.08  & 72.57\cr
  NGC4102  & SAB(s)b?;Liner  &  48.1  &  0.68  &  0.54  & 245.5\cr
  NGC4194  &  IBm;pec  & 23.81  &  0.95  &  0.63  &109\cr
  NGC4418  & (R\')SAB(s)a  & 43.89  &  1.37  &  1.15  &183\cr
  NGC4490  &  SB(s)d;pec  &  45.9  &0.6  & -0.21  & 245.6\cr
  NGC4519  &  SB(rs)d  &  3.74  &  0.53  &  -0.3  & 21.05\cr
  NGC4691  & (R)SB(s)0/a;pec  & 14.43  &  0.62  & -0.02  & 76.23\cr
  NGC4713  & SAB(rs)d  &4.6  &  0.42  &  -0.3  & 28.75\cr
IC3908  &  SB(s)d?  &  8.09  &  0.47  &  0.29  & 47.99\cr
IC0860  &  SB(s)a:  & 17.93  &  0.96  &  1.05  & 81.82\cr
IC0883  & Pec  & 17.01  &0.7  &  1.14  & 85.91\cr
  NGC5433  & SAB(s)c:  &  6.62  &  0.57  &  0.33  & 36.15\cr
  NGC5713  & SAB(rs)bc pec  & 21.89  &  0.57  &0.2  & 119.5\cr
  NGC5786  & (R\'\_2)SAB(s)bc  &  5.26  &  0.35  & -0.49  & 36.04\cr
  NGC5866  & S0\_3;HII/Liner  &  5.21  &0.3  &  -0.6  & 38.82\cr
 CGCG1510.8+0  & SB?(s?)0/a pec  & 20.84  &  0.66  &  1.56  & 107.5\cr
  NGC5962  & SA(r)C  &  8.89  &0.4  & -0.09  &  56.9\cr
IC4595  &SB?c sp II:  &  7.05  &  0.39  &  0.14  &  45.7\cr
  NGC6286  & SB(s)0+ pec?  &  8.22  &  0.37  &  0.83  & 54.71\cr
IC4662  &IBm  &  8.81  &  0.74  & -0.38  & 43.64\cr
  NGC6753  &(R)SA(r)b  &  9.77  &  0.35  & -0.02  & 66.93\cr
  NGC6821  &  SB(s)d:  &  3.63  &  0.64  & -0.21  & 18.95\cr
  NGC6958  & E+  &  1  &0.5  & -0.88  & 5.771\cr
  NGC7218  & SB(r)c  &  4.67  &  0.42  & -0.17  & 29.19\cr
  NGC7418  & SAB(rs)cd  &  5.38  &  0.33  & -0.32  & 38.03\cr
IC5325  &SAB(rs)bc  &  5.15  &  0.36  & -0.26  & 34.77\cr
 IRASF23365+3  & S?Ba? pec or Pec  &  7.44  &  0.84  &1.2  & 35.35\cr
  NGC7771  &SB(s)a  & 19.67  &  0.49  &  0.41  & 114.5\cr
  MRK0331  & SA(s)a: pec  & 18.04  &  0.76  &  1.12  & 88.55 \cr
\hline
\end{tabular}
\end{flushleft}
\end{table*}

\section{Measured Line fluxes}
\begin{table*}[htb]{}
\scriptsize
\caption[ ]{Line fluxes for the distant sample (in units of $10^{-14} W/m^2$)}
\begin{flushleft}
\begin{tabular}{lccccccl}
\hline
Galaxy & [CII](158$\mum$) & [OI](145$\mum$) & [NII](122$\mum$) & [OIII](88$\mum$) & [OI](63 $\mum$) & [NIII] (57$\mum$) & [OIII] (52$\mum$) \cr
\hline
 NGC0278   & 0.697   &  $<$   0.014   & 0.031   & 0.178	   & 0.312   &  $<$   0.093   &  $<$   0.102 \cr
  NGC520   & 0.254   & 0.025   & 0.049   &  0.13	   & 0.184   &  $<$   0.162   &  $<$   0.205 \cr
 NGC0693   & 0.167   &   -   &  $<$   0.011   &   -	   &  0.07   &   -   &   - \cr
 NGC0695   & 0.181   &   -   & 0.015   &  0.07	   & 0.109   &   -   &   - \cr
 UGC01449   & 0.143   &   -   &  $<$0.01   &   -	   & 0.081   &   -   &   - \cr
 MCG-03-06-01   & 0.035   &   -   &   -   &   -	   & 0.035   &   -   &   - \cr
 NGC0986   & 0.254   &  $<$   0.027   &  $<$   0.034   &  $<$   0.065	   & 0.103   &  $<$   0.135   &  $<$   0.068 \cr
 NGC1022   & 0.127   &   -   &  $<$   0.024   &  $<$   0.058	   & 0.191   &   -   &   - \cr
 NGC1052   & 0.012   &   -   &   -   &   -	   &   -   &   -   &   - \cr
 UGC02238   & 0.222   &   -   &  $<$   0.015   & 0.068	   & 0.073   &   -   &   - \cr
 NGC1155   & 0.034   &   -   &   -   &   -	   & 0.035   &   -   &   - \cr
 NGC1156   & 0.186   &   -   &  $<$0.01   &   -	   &  0.07   &   -   &   - \cr
 NGC1222   & 0.206   &   -   &  $<$   0.013   &   -	   &  0.25   &   -   &   - \cr
UGC02519   & 0.157   &   -   &   -   &   -	   &  $<$   0.059   &   -   &   - \cr
 NGC1266   & 0.038   &   -   &  $<$   0.016   &  $<$   0.035	   & 0.054   &   -   &   - \cr
 NGC1317   & 0.074   &   -   &   -   &   -	   & 0.041   &   -   &   - \cr
 NGC1326   & 0.148   &   -   &  $<$0.01   &   -	   & 0.076   &   -   &   - \cr
 NGC1385   & 0.511   &   -   &  $<$0.02   & 0.232	   & 0.243   &  $<$   0.071   &   - \cr
UGC02855   & 0.525   &  $<$   0.033   &  0.04   & 0.108	   & 0.247   &  $<$   0.164   &  $<$   0.198 \cr
 NGC1482   & 0.655   &  $<$   0.034   & 0.027   &  0.13	   & 0.318   & 0.055   &  $<$   0.094 \cr
 NGC1546   &  0.27   &   -   & 0.044   &   -	   & 0.062   &   -   &   - \cr
 NGC1569   & 0.674   &  $<$   0.014   &  $<$   0.028   & 2.663	   & 0.616   &  $<$   0.141   & 1.722 \cr
 NGC2388   & 0.191   &   -   & 0.045   & 0.035	   & 0.097   &  $<$   0.081   &   - \cr
 ESO317-G023   & 0.101   &   -   & 0.018   & 0.035	   & 0.068   &  $<$   0.071   &   - \cr
 IRASF10565+2   & 0.042   &   -   &  $<$   0.014   &  $<$   0.023	   & 0.076   &   -   &   - \cr
 NGC3583   & 0.147   &   -   &  $<$   0.018   &   -	   & 0.088   &   -   &   - \cr
 NGC3620   &  0.25   & 0.029   &  $<$   0.033   &  $<$0.06	   & 0.164   &  $<$   0.158   &  $<$   0.166 \cr
 NGC3683   & 0.376   &   -   &  $<$   0.014   & 0.127	   &  0.17   &  $<$   0.055   &   - \cr
 NGC3705   & 0.057   &   -   &  $<$   0.004   &   -	   &   -   &   -   &   - \cr
 NGC3885   & 0.137   &   -   & 0.025   & 0.032	   &  0.11   &   -   &   - \cr
 NGC3949   &  0.26   &   -   &  $<$   0.025   & 0.149	   & 0.183   &   -   &   - \cr
 NGC4027   & 0.287   &   -   &  $<$   0.015   & 0.125	   & 0.148   & 0.062   &   - \cr
 NGC4102   & 0.286   & 0.022   &  $<$   0.051   &  $<$   0.101	   & 0.268   &  $<$   0.232   &  $<$   0.147 \cr
 NGC4194   & 0.189   &   -   &  $<$   0.021   & 0.189	   & 0.243   & 0.062   &   - \cr
 NGC4418   &  $<$   0.028   &  $<$0.02   &  $<$   0.014   & 0.045	   &  $<$   0.053   &  $<$   0.214   &  $<$   0.213 \cr
 NGC4490   & 0.423   & 0.011   &  $<$   0.032   & 0.502	   & 0.328   &  $<$   0.229   & 0.403 \cr
 NGC4519   & 0.069   &   -   &  $<$   0.004   &   -	   &   -   &   -   &   - \cr
 NGC4691   & 0.222   &   -   &  $<$   0.037   & 0.129	   & 0.158   &  $<$   0.103   &   - \cr
 NGC4713   & 0.137   &   -   &  $<$   0.007   &   -	   & 0.085   &   -   &   - \cr
  IC3908   & 0.205   &   -   &  $<$   0.021   & 0.077	   & 0.115   &   -   &   - \cr
  IC0860   &  $<$   0.016   &   -   &  $<$   0.021   &  $<$   0.031	   &  $<$   0.025   &  $<$   0.346   &   - \cr
  IC0883   & 0.088   &   -   &  $<$   0.015   & 0.049	   &  0.12   &   -   &   - \cr
 NGC5433   & 0.157   &   -   &   -   &   -	   & 0.101   &   -   &   - \cr
 NGC5713   & 0.454   &  $<$   0.009   & 0.053   & 0.149	   & 0.265   &  $<$   0.024   & 0.143 \cr
 NGC5786   & 0.123   &   -   & 0.019   &   -	   &   -   &   -   &   - \cr
 NGC5866   & 0.052   &   -   & 0.012   &   -	   &  $<$   0.029   &   -   &   - \cr
 CGCG1510.8+0   &  0.04   &   -   &  $<$   0.025   &  $<$0.04	   &  $<$0.02   &  $<$   0.059   &   - \cr
 NGC5962   & 0.271   &   -   &  $<$   0.015   &   -	   & 0.083   &   -   &   - \cr
  IC4595   & 0.244   &   -   &  0.03   &   -	   & 0.069   &   -   &   - \cr
 NGC6286   & 0.169   &   -   &  $<$   0.012   & 0.018	   & 0.068   &   -   &   - \cr
  IC4662   & 0.094   &   -   &  $<$   0.009   & 0.401	   & 0.143   &   -   &   - \cr
 NGC6753   & 0.288   &   -   & 0.063   &  $<$0.03	   & 0.081   &   -   &   - \cr
 NGC6821   & 0.089   &   -   & 0.006   &   -	   &   -   &   -   &   - \cr
 NGC6958   &  0.008   &   -   &   -   &   -	   &  $<$   0.021   &   -   &   - \cr
 NGC7218   &  0.18   &   -   &  $<$0.01   &   -	   & 0.076   &   -   &   - \cr
 NGC7418   & 0.123   &   -   & 0.013   &   -	   &   -   &   -   &   - \cr
  IC5325   & 0.191   &   -   & 0.014   &   -	   &   -   &   -   &   - \cr
 IRASF23365+3   & 0.017   &   -   &   -   &   -	   &  $<$   0.023   &   -   &   - \cr
 NGC7771   & 0.307   &   -   & 0.053   & 0.056	   & 0.091   &  $<$   0.029   &   - \cr
 MRK0331   & 0.137   &   -   & 0.022   &  $<$   0.046	   & 0.107   &   -   &   -  \cr

\hline
\end{tabular}
\end{flushleft}
\end{table*}

\notetoeditor{Please place Table 6 here}
\section{Physical quantities derived from comparison with models}
\notetoeditor{Please place Table 7 here}
\begin{table*}[htb]{}
\scriptsize
\caption[ ]{Average physical parameters derived by comparing FIR line fluxes with PDR models}
\begin{flushleft}
\begin{tabular}{lccccccl}
\hline
\noalign{\smallskip}
Galaxy &  Log(G$_0$) &	Log(n) $\cc$ & Log(G$_0$) &	Log(n)	$\cc$ & Temperature & Pressure & Pressure (HII) \cr
       & (from & $\orat$)	   &  (from L$_{\rm [OI]}/\rm L_{\rm [CII]_c}$& and ($\rm L_{\rm [OI]}+\rm L_{\rm [CII]_c})/\rm L_{\rm TIR}$)   & K 	      & K$\cc$ & K$\cc$ \cr
\hline
     NGC0278 &         3.2 &	3.1	& 2.3 &        2.75 &         225 & $    1.3 \times 10^5$ & \cr
      NGC520 &        4.0 &	3.0	& 4.2 &         3.7 &         510 & $    2.6 \times 10^6$ & \cr
     NGC0693 &         3.3 &	3.1	& 2.7 &         2.6 &         310 & $ 1.2 \times 10^5$ & \cr
     NGC0695 &         3.4 &	3.3	& 3.2 &         3.2 &         340 & $    5.4 \times 10^5 $& \cr
    UGC01449 &        3.4 &	3.4	& 2.75 &           3  &        280 & $    2.8 \times 10^5 $& \cr
 MCG-03-06-01 &      4.3 &	4.1	& ...   & 	...  &		550 &$ 6.9\times 10^6 $& \cr
     NGC0986 &        3.0 &	2.1	&  2.8 &           2  &       600 &  $     6 \times 10^4 $& \cr
     NGC1022 &         3.6 &	3.4	& 3.8 &         3.6 &         430 & $    1.7 \times 10^6 $& \cr
    UGC02238 &        3.2 &	2.8 	&  2.9 &         2.3 &         460 & $    9.2 \times 10^4 $& \cr
     NGC1155 &      3.3	&	3.7	& ...	& 	... &		350 &$ 1.7 \times 10^6$& \cr
     NGC1156 &        3.1  &	3.0	&  2.6 &         2.6 &         290 & $    1.2 \times 10^5 $& \cr
     NGC1222 &         3.8 &	4.1	& 3.2 &         3.5 &         325 & $      1 \times 10^6 $& \cr
     NGC1266 &        3.7 &	3.0	& 4.9 &         4.2 &         900 & $    1.4 \times 10^7$ & \cr
    NGC1317 &        3.1	& 2.3	& ...	& ...		&  575 & $      1.1 \times 10^5$ & \cr
     NGC1326 &        3.4 &	3.0	& 2.9 &         2.7 &         340 & $    1.7 \times 10^5$ & \cr
     NGC1385 &         3.0  &	2.8	& 2.8 &         2.7 &         310 & $    1.6 \times 10^5 $& \cr
    UGC02855 &        2.9 &	2.3	& 3.3 &         2.5 &         560 & $    1.8 \times 10^5$ & \cr
     NGC1482 &        3.6 &	3.4	& 2.9 &         2.8 &         325 & $      2 \times 10^5$ & \cr 
     NGC1546 &         2.5	& 1.9	& 3.3 &         2.6 &         525 & $    2.1 \times 10^5 $& \cr
     NGC1569 &        4.0	& 3.9	& 3.1 &         3.2 &         325 & $    5.2  \times 10^5 $& $2.5 \times 10^5$\cr
     NGC2388 &        3.6	& 3.0	& 4.7 &         4.2 &         775 & $    1.2  \times 10^7$ & \cr
 ESO317-G023 &        3.3	& 2.6	& 4.2   &      3.4    &     550  &  $ 1.4  \times 10^6    $& \cr
 IRASF10565+2&        3.7	& 3.25	& 4.4 &        3.9  &       580   & $ 4.6  \times 10^6 $& \cr
     NGC3583 &         2.7	& 2.5	& 3.5 &           3 &        440  & $   4.4  \times 10^5 $& \cr
     NGC3620 &        3.6	& 2.7	& 3.6 &         2.7 &         600 & $      3  \times 10^5 $& \cr
     NGC3683 &         2.9	& 2.7	& 2.7 &        2.65 &         300 & $    1.3  \times 10^5 $& \cr
     NGC3885 &         3.6	& 3.4	& 3.9 &         3.6 &         450 & $    1.8  \times 10^6 $& \cr
     NGC3949 &         2.9	& 2.9	& 3.1 &           3  &       340 &  $   3.5  \times 10^5 $& \cr
     NGC4027 &        2.8	& 2.6	& 2.85 &         2.7 &         325 & $    1.6  \times 10^5 $& \cr
     NGC4102 &        3.5	& 3.0	& 4.8 &         3.2 &         900 & $     1.4  \times 10^6 $ & \cr
     NGC4194 &         4.0	& 3.7	& 3.4 &         3.4 &         370 & $    9.3  \times 10^5 $& \cr
     NGC4490 &        3.4	& 2.9	& 3.45 &         2.9 &         450 & $    3.6  \times 10^5 $& $8 \times 10^5$\cr
     NGC4691 &         3.4	& 3.3	& 3.3 &         3.2 &         360 & $    5.7  \times 10^5 $& \cr
     NGC4713 &         2.8	& 2.8	& 2.9 &         2.9 &         310 & $    2.5  \times 10^5 $& \cr
      IC3908 &         3.0	& 2.9	&   3 &         2.9 &        340 &  $   2.7  \times 10^5 $& \cr
      IC0883 &        3.6	& 3.2	& 3.95 &        3.45 &         475 & $    1.3  \times 10^6 $& \cr
     NGC5433 &      3.8		& 3.3	& ...  & ...         &  450    & $ 8.9 \times 10^5 $& \cr
     NGC5713 &        3.3	& 3.2	& 3.25 &         3.2 &         350 & $    5.5 \times 10^5 $& $1.3 \times 10^6$\cr
     NGC5962 &         2.8	& 2.4	& 2.6 &        2.25 &         360 & $    6.4 \times 10^4 $& \cr
      IC4595 &         2.7	& 2.4	& 2.9 &         2.5 &         390 & $    1.2 \times 10^5 $& \cr
      IC4662 &        3.6	& 3.7	& 3.4 &         3.5 &         360 & $    1.1 \times 10^6 $& \cr
     NGC6286 &         2.7	& 2.2	& 2.9 &         2.4 &         425 & $    1.1 \times 10^5 $& \cr
     NGC6753 &       2.6	& 2.2	& 4.1 &         2.3 &        1400 &$     2.8 \times 10^5 $& \cr
     NGC7218 &         2.8	& 2.7	& 2.6 &         2.7 &         270 & $    1.4 \times 10^5 $& \cr
     NGC7771 &         3.0	& 2.2	& 3.6 &        2.75 &         560 & $    3.1 \times 10^5 $& \cr
     MRK0331 &        3.7	& 3.2	& 3.9 &         3.4 &         460 & $    1.2 \times 10^6 $& \cr

\noalign{\smallskip}
\hline
\end{tabular}
\end{flushleft}
\end{table*}

\end{document}